\documentclass{article}

\usepackage{arxiv}

\usepackage[utf8]{inputenc} % allow utf-8 input
\usepackage[T1]{fontenc}    % use 8-bit T1 fonts
\usepackage{hyperref}       % hyperlinks
\usepackage{url}            % simple URL typesetting
\usepackage{booktabs}       % professional-quality tables
\usepackage{array}
\usepackage{amsfonts}       % blackboard math symbols
\usepackage{nicefrac}       % compact symbols for 1/2, etc.
\usepackage{graphicx}
\usepackage{amssymb} % for \coloneqq
\usepackage{subcaption}% for subfigs
\usepackage{adjustbox}
\usepackage[numbers]{natbib}
\usepackage{doi}
\usepackage{amsmath}
\usepackage{amssymb}
\usepackage{algorithm}
\usepackage{algpseudocode}
\usepackage{xcolor}
\usepackage{fancyhdr}
\usepackage{amsthm}
\usepackage{appendix} 
\usepackage{cleveref}
\usepackage{bm}
\usepackage[mathscr]{euscript}
\usepackage{textcomp}
\usepackage{tcolorbox}

\DeclareMathAlphabet{\pazocal}{OMS}{zplm}{m}{n}

\definecolor{Rone}{RGB}{0,0,0}         % Black
\definecolor{Rtwo}{RGB}{0,0,0}         % Black

\usepackage{lscape}		
\usepackage{fancyvrb}

\newtheoremstyle{remarkstyle}  % Name
  {5pt}                        % Space above
  {5pt}                        % Space below
  {}                           % Body font (upright)
  {}                           % Indent amount
  {\bfseries}                  % Theorem head font (bold)
  {.}                          % Punctuation after theorem head
  { }                          % Space after theorem head
  {}                           % Theorem head spec (empty = `normal`)
\newtheorem{theorem}{Theorem}[section]

\theoremstyle{remarkstyle}
\newtheorem{remark}[theorem]{Remark}

\crefname{figure}{Fig.}{Figs.}
\Crefname{figure}{Figure}{Figures}
\crefname{algorithm}{Algorithm}{Algorithms}
\crefname{equation}{Eq.}{Eqs.}
\crefname{section}{Section}{Sections}
\crefname{appendix}{Appendix}{Appendices}
\crefname{section}{Appendix}{Appendices}

\newcommand{\alert}[1]{\textcolor{black}{#1}}

\title{
Conformal Quantile Regression for Neural Probabilistic Constitutive Modeling
}

\author{
Bahador Bahmani\\
Department of Mechanical Engineering, Northwestern University\\
Evanston, IL, USA
}

\author{
Bahador Bahmani\\
Department of Mechanical Engineering, Northwestern University\\
Evanston, IL, USA\\
\texttt{bahador.bahmani@northwestern.edu}
}

\renewcommand{\shorttitle}{CQR for Anisotropic Soft Materials}

\date{}

\hypersetup{
    colorlinks=true,      % Enable colored links
    linkcolor=green,      % Color for internal links
    urlcolor=blue,        % Color for URLs
    citecolor=blue        % Color for citations
}

\begin{document}
\maketitle

\begin{abstract}
    Biological soft tissues exhibit substantial inter-subject variability, making the automation of constitutive material modeling essential for patient-specific analysis and design. Such materials are not only highly nonlinear but also display intrinsic stochasticity arising from their complex and heterogeneous microstructure. Despite recent advances in data-driven constitutive modeling, most existing approaches remain deterministic and fail to quantify predictive uncertainty, thereby limiting their reliability in downstream mechanical analyses. In this work, we propose a probabilistic, data-driven constitutive modeling framework for anisotropic soft materials that explicitly accounts for uncertainty through conformalized quantile regression applied to tensor-valued fields.
    %We further extend conformal quantile regression to tensor-valued constitutive responses in the context of hyperelastic material modeling. 
    The proposed framework is built upon a strain-invariant, polyconvex formulation that ensures thermodynamic consistency and promotes robust predictive performance, including in extrapolative regimes. A key advantage of the proposed approach is its simplicity: it can be applied in a \textit{plug-and-play} manner to endow existing deterministic models with probabilistic predictions, while remaining distribution-free and requiring no assumptions on the underlying data distribution. Moreover, the method is straightforward to train, scalable to models with a large number of parameters, and avoids Monte Carlo sampling at inference, making it computationally efficient and well suited for uncertainty propagation in large-scale mechanical simulations. The proposed method is validated using several benchmark datasets synthesized and collected from the literature.
\end{abstract}

% keywords can be removed

\section{Introduction}

Biological soft tissues exhibit substantial inter-subject variability, making the automation of constitutive modeling essential for patient-specific analysis and design \cite{arnold2023review,budday2019fifty,jacquet2017intra}. Such materials are not only highly nonlinear but also display intrinsic stochasticity arising from their complex and heterogeneous microstructure \cite{alzweighi2021influence,mihai2018stochastic}. Despite recent advances in data-driven constitutive modeling \cite{fuhg2024review}, most existing approaches remain deterministic and fail to quantify predictive uncertainty, thereby limiting their reliability in downstream mechanical analyses.

Neural network–based constitutive models have come to dominate recent developments in data-driven constitutive modeling \cite{mozaffar2019deep,teichert2019machine,vlassis2020geometric,linka2021constitutive,fuhg2022learning,tac2022data,thakolkaran2022nn,linden2023neural,chen2022polyconvex,bahmani2024physics,martonova2025generalized}, owing to their expressive power and flexibility. When appropriately constrained by thermodynamic principles—such as material frame indifference, energy consistency—these models can provide robust and physically admissible representations of material behavior.

In this work, we construct a physics-encoded, anisotropic data-driven constitutive model, building on the seminal framework of \cite{schroder2008anisotropic}. Anisotropy is incorporated through structural tensors that encode the material symmetries associated with the underlying anisotropic directions.%Our formulation treats them as part of the identification process, allowing the anisotropic structure to be inferred directly from data.
{\color{Rone}{
%The strain energy is represented in an additively separable invariant-based form, corresponding to a structured subclass of polyconvex energy functionals. This modeling choice enables monotone univariate parameterizations and tractable enforcement of polyconvexity, while also facilitating post-hoc interpretability analyses by isolating the contributions of individual invariants, which may be advantageous in certain applications. The trade-off is that explicit cross-invariant coupling terms are not represented, which may result in a suboptimal model for material systems that require such interactions.
%
%
%The strain energy is represented in an additively separable invariant-based form, corresponding to a structured \textit{subclass} of polyconvex energy functionals. Such separable representations have been successfully employed in a broad range of constitutive models for anisotropic materials \cite{schroder2003invariant,schroder2008anisotropic,ehret2007polyconvex,balzani2008analysis,linka2021constitutive,linden2023neural}. This modeling choice enables monotone univariate parameterizations and tractable enforcement of polyconvexity. While explicit cross-invariant coupling terms are not represented, the resulting structure can facilitate data-efficient learning when the separability assumption is consistent with the underlying material behavior.
The strain energy is represented in an \textit{additively separable} invariant-based form, corresponding to a structured \textit{subclass} of polyconvex energy functionals.}}
Within this formulation, we consider neural network parameterizations for the phenomenological components of the constitutive response that preserve polyconvexity by construction. Rather than parameterizing the unknown strain energy density itself \cite{kalina2025neural,bahmani2024physics}, we parameterize its required gradients, which enter directly into the  stress response. This choice leads to improved trainability and reduced computational cost.

Extending neural network–based constitutive models to a probabilistic setting introduces additional challenges. One promising direction is to formulate  parametric models within a Bayesian framework, which has proven effective for constitutive models with a relatively small number of unknown parameters \cite{doraiswamy2013bayesian,seyedsalehi2015prior,ritto2015bayesian,madireddy2016bayesian,akintunde2019bayesian,joshi2022bayesian}. However, a direct Bayesian treatment that places prior distributions over all network parameters is often computationally prohibitive \cite{izmailov2021bayesian}, as even moderately sized neural networks involve thousands to hundreds of thousands of parameters. The resulting posterior distributions are high-dimensional, strongly coupled, and highly non-Gaussian, rendering analytical inference intractable. Consequently, training typically relies on approximate inference techniques, such as variational inference, which may suffer from biased uncertainty estimates and underestimation of posterior variance due to restrictive variational families \cite{blei2017variational}. While asymptotically exact Markov chain Monte Carlo (MCMC) methods—such as Hamiltonian Monte Carlo (HMC)—can, in principle, approximate the full posterior over network parameters, their computational cost scales poorly with model dimension and dataset size \cite{neal2012bayesian}. 

Moreover, uncertainty quantification in Bayesian neural networks typically relies on Monte Carlo estimation of the posterior predictive distribution, requiring multiple forward passes at inference. This becomes especially burdensome for uncertainty propagation in large-scale mechanical simulations, where each forward evaluation may already be computationally expensive and repeated sampling is prohibitive.

%The enforcement of thermodynamic constraints in a fully probabilistic setting presents an additional challenge. Incorporating physical principles—such as polyconvexity, material stability—into Bayesian neural network frameworks \cite{blundell2015weight} is generally nontrivial and may require specialized parameterizations or carefully designed constrained priors to ensure that both prior and posterior distributions remain physically admissible \cite{brewick2018uncertainty}.

{\color{Rone}{The enforcement of thermodynamic constraints in a fully Bayesian setting \cite{brewick2018uncertainty,blundell2015weight} may require additional care. When physical admissibility is ensured intrinsically through architectural design (e.g., convex or polyconvex neural parameterizations), these structural properties can, in principle, be preserved under Bayesian inference. However, even in such intrinsically constrained formulations, the \textit{prior distribution} must have support entirely within the admissible parameter space. For example, in convex neural networks certain weights are required to remain nonnegative to guarantee convexity; in a Bayesian formulation, this restricts the class of admissible priors to distributions with positive support (e.g., truncated Gaussian or log-normal distributions). Such restrictions can alter posterior geometry and complicate inference, particularly for variational approximations.
}}
%When constraint satisfaction instead relies on loss penalties or parameter bounds, maintaining admissibility under posterior sampling becomes even more challenging, as both prior and posterior distributions must remain within the physically consistent set.

In light of these challenges, we adopt a different strategy while still leveraging the expressive power of neural network parameterizations. Rather than pursuing a fully Bayesian treatment, we draw on ideas from quantile regression \cite{koenker2001quantile,blum2024use,chen2026uncertainty}, a classical \textit{frequentist} approach to probabilistic modeling that enables uncertainty quantification by directly learning conditional quantiles of the response. This formulation allows deterministic constitutive models to be extended to probabilistic settings without introducing the additional complexity associated with Bayesian inference over high-dimensional parameter spaces.

Crucially, because uncertainty is modeled at the output level rather than through distributions over network parameters, quantile regression permits thermodynamic constraints to be enforced pointwise for each quantile realization using the same physics-constrained parameterizations as in the deterministic setting. This property makes it straightforward to preserve physical admissibility—such as material stability, dissipation, and polyconvexity—even in the presence of uncertainty.

While quantile regression is conceptually simple, it does not assume a specific parametric form for the conditional predictive distribution. This property is particularly advantageous in applications such as constitutive material modeling, where strong nonlinearities and complex material behavior make the underlying conditional distribution unknown \textit{a priori}. As a result, distribution-free methods such as quantile regression are both highly desirable and broadly applicable for uncertainty quantification in nonlinear mechanical systems.

%In this work, we construct a physics-encoded, anisotropic data-driven constitutive model, building on the seminal framework of \cite{schroder2008anisotropic}. Anisotropy is incorporated through structural tensors that encode the material symmetries associated with the underlying anisotropic directions. Our formulation treats them as part of the identification process, allowing the anisotropic structure to be inferred directly from data.

%Within this formulation, we consider neural network parameterizations for the phenomenological components of the constitutive response, designed to preserve polyconvexity by construction. Rather than parameterizing the unknown strain energy density itself, we parameterize its required gradients, which enter directly into the constitutive stress response. This choice leads to improved trainability and reduced computational cost.

We extend the proposed deterministic data-driven constitutive model to a probabilistic setting by replacing the unknown functions with their corresponding quantile representations and learning these quantile functions directly from data. Specifically, the model is trained such that the predicted stress quantiles match the empirical conditional stress quantiles observed in the data. To address potential miscalibration of quantile-based uncertainty estimates arising from finite data, we conformalize our quantile regression formulation following \cite{romano2019conformalized}. In particular, we extend this formulation to tensor-valued constitutive responses, enabling calibrated uncertainty quantification for stress fields. 

{\color{Rone}
\remark{
In this work, our primary focus is not on strictly real-time prediction, but rather on settings where data are sparse and noisy, and reliable uncertainty quantification is essential for decision-making and robust design under uncertainty. While probabilistic extensions of deterministic models often introduce additional computational cost, the proposed quantile-based approach avoids sampling during both training and inference. As a result, its computational cost at inference is comparable to that of a deterministic model.
}
}

\subsection*{Other Related Work}
Mihai et al.~\cite{mihai2018stochastic} developed stochastic hyperelastic models by treating constitutive parameters as random variables calibrated from experimentally measured mean values and standard deviations, enabling the propagation of material variability through continuum models. 
This approach is primarily limited to low-order statistical moments and has been demonstrated mainly for isotropic material responses.
%Early probabilistic constitutive modeling efforts focused on Bayesian calibration of classical material models. 
%
%Brewick et al.~\cite{brewick2018uncertainty} introduced a physics-constrained Bayesian framework for uncertainty quantification and parameter identification of classical hyperelastic material models, employing MCMC sampling to infer posterior distributions over material parameters. 
Full Bayesian treatments of model parameters in classical constitutive models have been studied in several works \cite{brewick2018uncertainty,doraiswamy2013bayesian,seyedsalehi2015prior,ritto2015bayesian,madireddy2016bayesian,akintunde2019bayesian,joshi2022bayesian}.
While statistically rigorous, such full Bayesian treatments can become computationally demanding as the dimensionality of the parameter space increases. 
%Fuhg et al.~\cite{fuhg2022physics} proposed a framework based on irreducible integrity basis representations, where stress coefficients are modeled as Gaussian process functions of strain invariants. 
Gaussian process regression has also been leveraged for soft material modeling \cite{frankel2020tensor,fuhg2022physics}. While Gaussian processes provide principled uncertainty estimates, their reliance on conditional Gaussianity may be restrictive for highly nonlinear constitutive responses, and their computational cost can further limit scalability to high-dimensional or large-scale datasets.

More recently, Linka et al.~\cite{linka2025discovering} proposed a probabilistic calibration framework for constitutive network models \cite{linka2023new} of isotropic soft materials by introducing probability distributions over unknown parameters and training the resulting model using variational inference.
As with other variational Bayesian approaches, the quality of the inferred uncertainty depends on the expressiveness of the variational family and may suffer from biased posterior approximations. McCulloch and Kuhl~\cite{mcculloch2025discovering} introduced Gaussian constitutive neural networks for anisotropic soft materials, in which neural networks parameterize the conditional mean and variance of the stress response under an assumed Gaussian likelihood. 
While effective, this formulation relies on conditional Gaussianity assumptions, and enforcing thermodynamic constraints on the predicted variance fields in a physically consistent manner is not straightforward.

%Finally, Wollner et al.~\cite{wollner2025reparameterization} developed a reparameterization-invariant Bayesian framework for probabilistic calibration of simple constitutive models, deriving parameterization-independent posteriors and employing MCMC-based inference for uncertainty estimation.

\subsection*{\textcolor{Rtwo}{Paper Organization}}
The remainder of the paper is organized as follows. In Section~\ref{sec:formulation-all}, we present the proposed formulation, beginning with the deterministic constitutive model and subsequently extending it to a probabilistic setting via quantile regression, followed by conformalized quantile regression for calibrated uncertainty quantification. In Section~\ref{sec:examples}, we demonstrate the performance of the proposed method through three numerical examples. Finally, Section~\ref{sec:conclud} concludes the paper with a summary of the main findings and a discussion of the limitations of the proposed approach.

\section{Formulation}
\label{sec:formulation-all}
In this section, we first describe how the proposed formulation enforces physical constraints by construction for anisotropic materials. We then motivate the parameterization of the unknown phenomenological components of the general constitutive framework and discuss the resulting benefits in terms of robustness of training and computational efficiency. The deterministic formulation is subsequently extended to a probabilistic setting through quantile regression. Finally, the framework is further extended to a conformalized quantile regression formulation for tensor-valued fields to obtain calibrated uncertainty estimates for constitutive responses.

\subsection{Deterministic Constitutive Backbone}
We begin by introducing the deterministic constitutive framework that forms the backbone of the proposed probabilistic model. The formulation is based on seminal work of~\cite{schroder2008anisotropic} and is designed to provide a physically consistent, flexible, and expressive representation of anisotropic hyperelastic materials. This deterministic model serves as a structural prior that enforces essential mechanical principles—such as material symmetry, polyconvexity, and thermodynamic consistency—while remaining amenable to data-driven calibration.

\textcolor{Rtwo}{We restrict attention to incompressible materials, which is a common and appropriate assumption for many soft biological and polymeric solids. Under this assumption, the third isotropic invariant is constant and therefore does not enter the strain energy representation explicitly.}
Specifically, the strain energy density is expressed as
\begin{equation}
    \psi(\boldsymbol{F}; \{\overset{j}{\boldsymbol{G}}\}_{j=1}^{N_G})
    = \sum_{i=1}^2 f_i(I_i)
    + \sum_{j=1}^{N_G} [ \bar{f}_{4j}(I_{4j}) + \hat{f}_{5j}(I_{5j}) ],
\label{eq:neff}
\end{equation}
where $\boldsymbol{F}$ is the deformation gradient tensor, and $I_i$ are the isotropic strain invariants of the right Cauchy--Green deformation tensor $\boldsymbol{C} = \boldsymbol{F}^T \boldsymbol{F}$, defined as $I_1 = \mathrm{tr}(\boldsymbol{C})$ and $I_2 = \mathrm{tr}(\mathrm{adj}(\boldsymbol{C}))$. The adjugate and cofactor operators satisfy $\mathrm{adj}(\boldsymbol{C}) = \mathrm{cof}(\boldsymbol{C})^T = \det(\boldsymbol{C})\,\boldsymbol{C}^{-1}$.

The second-order, positive semi-definite structural tensors $\overset{j}{\boldsymbol{G}}$ are introduced to capture anisotropy arising from possible fiber orientations within the continuum. They introduce additional degrees of freedom through the anisotropic strain invariants
\begin{align}
    I_{4j} = \mathrm{tr}\![\boldsymbol{C}\,\overset{j}{\boldsymbol{G}}],\quad I_{5j} = \mathrm{tr}\![\mathrm{cof}(\boldsymbol{C})\,\overset{j}{\boldsymbol{G}}].
\end{align}
The above strain energy functional is employed for incompressible materials; a physically justified simplification for most biological tissues~\cite{holzapfel2002nonlinear,ogden1997non}. Incompressibility is enforced by introducing a constraint term in the total strain energy, expressed as $p\,(J-1)$, where $J = \det(\boldsymbol{F})$ and $p$ is an additional unknown scalar field that enforces the required hydrostatic pressure to satisfy the incompressibility condition $J \approx 1$.

The functions $f_i$, $\bar{f}_{4j}$, and $\hat{f}_{5j}$ are \textit{unknown} real-valued scalar functions. 
If the scalar functions are chosen to be convex with respect to their arguments, the resulting strain energy density is polyconvex in $\boldsymbol{F}$~\cite{marsden1994mathematical,ball1976convexity}. As a consequence, the associated variational problem admits minimizers under standard boundary conditions, and the resulting constitutive model satisfies rank-one convexity, which is closely related to mechanical stability. 
Moreover, adherence to this framework guarantees thermodynamic consistency \textcolor{Rtwo}{(the Clausius--Duhem inequality is satisfied with identically zero dissipation)}, and satisfies the required symmetry properties with respect to the structural tensors (or fiber orientations).

With a mild abuse of notation, the strain energy density can be expressed as a sum of univariate basis functions of a collection of strain invariants $\tilde{I}_k$, which collectively enumerate both the isotropic and anisotropic invariants introduced in~\cref{eq:neff}:
\begin{equation}
    \psi\!(\boldsymbol{F}; \{\overset{j}{\boldsymbol{G}}\}_{j=1}^{N_G})
    =
    \sum_{k=1}^{M}
    \tilde{f}_{k}(\tilde{I}_k),
    \qquad
    M = 2 + 2N_G .
    \label{eq:nergy-func}
\end{equation}
Each scalar function $\tilde{f}_{k}$ is required to be convex and non-decreasing with respect to its argument.

The first Piola--Kirchhoff stress tensor $\boldsymbol{P}$ is obtained by
\begin{equation}
    \boldsymbol{P}
    =
    \frac{\partial \psi}{\partial \boldsymbol{F}}
    =
    \sum_{k=1}^{M}
    \frac{\partial \tilde{I}_{k}}{\partial \boldsymbol{F}}
    \frac{d \tilde{f}_k}{d \tilde{I}_{k}}
    =
    \sum_{k=1}^{M}
    \frac{\partial \tilde{I}_{k}}{\partial \boldsymbol{F}}
    \tilde{g}_k(\tilde{I}_k),
\label{eq:stress-form}
\end{equation}
where $\tilde{g}_k(\tilde{I}_k) \equiv d\tilde{f}_k / d\tilde{I}_{k}$ denotes the derivative of the corresponding basis function. Since each $\tilde{f}_k$ is convex and non-decreasing, its derivative $\tilde{g}_k$ is \textbf{non-negative and monotone}, a property that plays a key role in the subsequent parameterization and learning strategy. For notational simplicity, the incompressibility contribution, which enters as an additional additive term $p\,\boldsymbol{F}^{-T}$, is not shown explicitly but is included in all computations.

{\color{Rone}{
\remark{
The representation in Equation~\ref{eq:neff} adopts an additively separable invariant-based structure, corresponding to a structured \textit{subclass} of polyconvex energy functionals. Additive invariant-based forms are widely used in anisotropic constitutive modeling \cite{holzapfel2000new,schroder2003invariant,schroder2008anisotropic,ehret2007polyconvex,balzani2008analysis,linka2021constitutive,linden2023neural}. While this assumption excludes explicit cross-invariant coupling terms, it enables monotone univariate parameterizations and tractable enforcement of polyconvexity. When appropriate for the material class under consideration, the resulting structured (low-rank) representation can improve data efficiency, reduce model complexity, and facilitate post-hoc interpretability and symbolic model extraction \cite{bahmani2024physics}.}}}

\textcolor{Rone}{
\remark{We note that representing anisotropy through fixed structure tensors is convenient but may be restrictive when the underlying fibre architecture is spatially varying or better described by an orientation distribution. Related distributed-orientation descriptions have been considered in earlier work by Pinsky and co-workers \citep{pinsky2005computational,cheng2015structural}.}
}

\subsection{Parameterization Strategy}

Two natural parameterization strategies arise for representing these basis functions.
In the first approach, the strain energy contribution is parameterized directly using a convex function
$\tilde{f}_k(x;\theta)$, for example through a convex neural network. The corresponding stress contribution
is then obtained by differentiation,
\begin{equation}
\tilde{g}_k(x;\theta) = \frac{d\,\tilde{f}_k(x;\theta)}{d x}.
\end{equation}
Parameterizing convex functions of strain invariants has predominantly relied on input convex neural networks (ICNNs)~\cite{klein2022polyconvex,chen2022polyconvex,as2022mechanics,fuhg2022learning,linden2023neural,bahmani2024physics,kalina2025neural}. ICNNs are a class of neural network architectures that enforce convexity with respect to their inputs by construction~\cite{amos2017input}.

In the second approach, the derivative of the strain energy density is parameterized directly as a monotone
function $\tilde{g}_k(x;\theta)$. When an explicit energy representation is required, the corresponding
convex function can be recovered via integration,
\begin{equation}
\tilde{f}_k(x;\theta) = \int_{x_0}^{x} \tilde{g}_k(\xi;\theta)\, d\xi,
\end{equation}
where $x_0$ is an arbitrary reference point. The integration constant does not affect the resulting stress
response and is therefore inconsequential for constitutive calibration.

%Intuitively, searching over monotone functions is simpler and more stable than searching over convex functions. 
Once $\tilde{g}_{k}$ has been learned, the corresponding energy functions $\tilde{f}_{k}$ can be recovered via integration, which is straightforward given the univariate nature of the invariant-based representation. In the following, we present a simple illustrative example that highlights the advantages of gradient-based parameterization over direct function-based approaches.

\subsubsection{Monotone Univariate Networks}
To parameterize the invariant-dependent derivative functions
$\tilde{g}_k(\cdot)$ introduced in the previous section, we employ a class
of monotone neural networks that guarantee non-negativity and monotonicity
of the learned mapping by construction \cite{sill1997monotonic}.

Each function $\tilde{g}_k$ is represented using a univariate feedforward
neural network of the form
\begin{equation}
\tilde{g}_k(x)
=
\mathcal{N}_k(x;\theta_k),
\end{equation}
where $\mathcal{N}_k:\mathbb{R}\to\mathbb{R}$ is constructed to be monotone
non-decreasing with respect to its input. Monotonicity is enforced by
restricting all network weights to be non-negative and by employing
activation functions with non-negative derivatives. Specifically, for a network with $L$ layers, the mapping takes the form
\begin{equation}
h^{(\ell+1)} = \sigma\!( W^{(\ell)} h^{(\ell)} + b^{(\ell)} ),
\qquad \ell = 0,\dots,L-1,
\end{equation}
with $h^{(0)} = x$, where $\sigma(\cdot)$ is the activation
function and the weight matrices satisfy
$
W^{(\ell)} = (\widehat{W}^{(\ell)})^2,
$
for unconstrained matrices $\widehat{W}^{(\ell)}$. This construction
guarantees elementwise non-negativity of the weights and, combined with the
choice of activation function, ensures that $\mathcal{N}_k$ is monotone
non-decreasing by design. A proof of the monotonicity of this neural network parameterization is provided in~\cref{app:proof_monotone_net}.

\paragraph{Numerical Illustration.}
%\label{sec:toy-monoton-ex}
This numerical illustration examines how the choice of parameterization affects optimization difficulty. Although the two formulations are mathematically equivalent at the functional level, they may induce different loss landscapes in parameter space, leading to distinct numerical performance during training.

\begin{figure}[h]
  \centering
  \begin{subfigure}[b]{0.33\textwidth}
    \includegraphics[width=\textwidth]{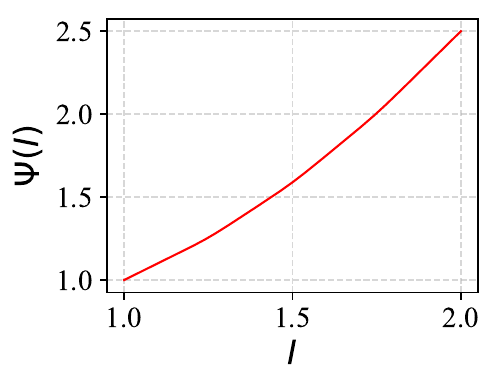}
    %\caption{}
    %\label{subfig:one}
  \end{subfigure}
    \begin{subfigure}[b]{0.33\textwidth}
    \includegraphics[width=\textwidth]{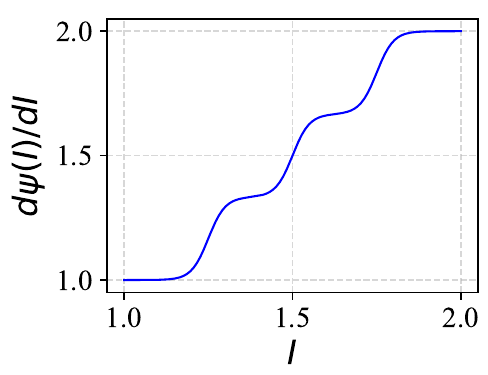}
    %\caption{}
    %\label{subfig:one}
  \end{subfigure}
  \caption{
  Synthetic example illustrating a smooth convex function (left) and its corresponding monotone derivative (right), used to compare performances of convex-function and gradient-based monotone parameterizations.
  }
  \label{fig:mono-vs-convex}
\end{figure}

We consider a univariate convex function, shown in the left panel of~\cref{fig:mono-vs-convex}, whose derivative exhibits a stepwise structure, shown in the right panel (see \cref{appx:monoton} for details). A total of 200 samples are drawn from the corresponding gradient field, and two regression settings are considered. In the first setting, the convex function is parameterized using a convex neural network and trained such that its gradient matches the sampled data. In the second setting, the gradient is parameterized directly using a monotone neural network and trained to fit the sampled gradient observations.

\begin{figure}[h]
  \centering
  \begin{subfigure}[b]{0.33\textwidth}
    \includegraphics[width=\textwidth]{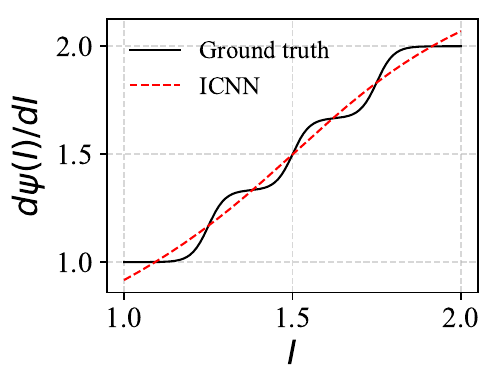}
    %\caption{}
    %\label{subfig:one}
  \end{subfigure}
    \begin{subfigure}[b]{0.33\textwidth}
    \includegraphics[width=\textwidth]{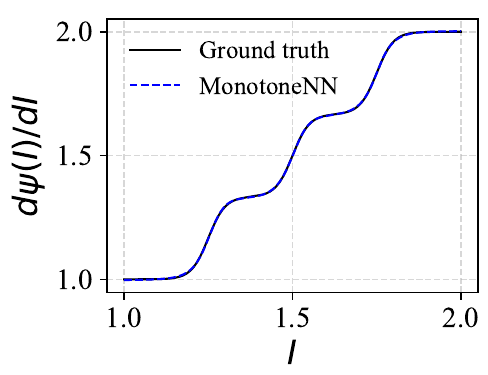}
    %\caption{}
    %\label{subfig:one}
  \end{subfigure}
  \caption{
  Predicted gradients obtained using convex energy parameterization (left) and gradient-based monotone parameterization (right). The monotone approach more accurately captures the underlying stepwise gradient structure in this example.
  }
  \label{fig:1d-predict-grad}
\end{figure}

Using the same number of trainable parameters and identical optimizer settings, the predicted gradients for the two modeling settings are shown in~\cref{fig:1d-predict-grad}. The results indicate that the monotone parameterization achieves higher accuracy than the convex parameterization in fitting the sampled gradient data. This behavior is consistent across different random initializations, suggesting that the monotone formulation provides a more effective parameterization for capturing the underlying gradient structure in this example. An additional advantage of the monotone approach is improved computational efficiency during both training and inference. Because the gradient is computed directly rather than via automatic differentiation, the monotone parameterization is observed to be approximately twice as fast (see~\cref{appx:monoton}).

\begin{remark}[\textbf{Other choices beyond neural networks}]
In this formulation, the basis functions are one-dimensional mappings, and thus a variety of function classes can be used in place of neural networks. Polynomials (in the global sense) are the simplest option, and there exist well-established methods for polynomial approximation under monotonicity constraints~\cite{shisha1965monotone,murray2016fast,mckay2011variable,curmei2025shape}. However, when the underlying function exhibits multi-scale structure, as in the pedagogical example above, polynomial representations may struggle to capture sharp local variations without resorting to high polynomial degrees. Such high-degree approximations are prone to oscillatory behavior and often exhibit poor extrapolation properties.
Spline-based representations provide another natural alternative, and monotonicity/convexity can be enforced through shape-restricted regression splines or constrained smoothing splines~\cite{ramsay1988monotone,meyer2008inference}. In addition, monotone piecewise-cubic interpolants have been studied extensively in numerical analysis~\cite{fritsch1980monotone}. A key practical limitation of spline models, however, is that their behavior outside the data range is largely determined by boundary conditions and knot placement, and is therefore often unreliable in extrapolative regimes. In mechanics applications, constitutive laws must remain robust under loading conditions that may fall outside the calibration domain, which motivates the neural parameterizations adopted in this work.
\end{remark}

\begin{remark}
    In this work, the enforcement of zero stress at the reference configuration is hard-coded using a scheme similar to that proposed in \cite{as2022mechanics,bahmani2024physics} and many other works.
\end{remark}

\subsection{A Probabilistic Constitutive Model via Tensorial Quantile Regression}
\label{sec:CQE}

In a data-driven setting, experimental variability, inter-subject heterogeneity, and measurement noise induce
stochasticity in the observed constitutive response. We therefore model the first Piola–Kirchhoff stress as a
tensor-valued random variable associated with the applied deformation, and assume that observed deformation--stress
pairs are generated according to an unknown joint distribution,
\begin{equation}
(\boldsymbol{F}, \boldsymbol{P}) \sim p_{\mathrm{data}},
\end{equation}
where $\boldsymbol{F}$ denotes the deformation gradient imposed by the experimental protocol and
$\boldsymbol{P}$ denotes the corresponding measured stress response.
The pairs $(\boldsymbol{F}, \boldsymbol{P})$ are assumed to be \textit{exchangeable}, which is sufficient for the validity of
the conformal guarantees derived later.
The dataset is partitioned into disjoint training $\mathcal{D}_{\mathrm{trn}}$, calibration
$\mathcal{D}_{\mathrm{cal}}$, and testing $\mathcal{D}_{\mathrm{tst}}$ subsets.

Our objective is to construct, for each deformation gradient $\boldsymbol{F}$, a set-valued predictor
$\mathcal{U}(\boldsymbol{F}) \subset \mathbb{R}^{3\times 3}$ such that the true stress lies in this set with high
probability.
To enable componentwise control of predictive uncertainty, we introduce a tensor of miscoverage levels
\begin{equation}
\boldsymbol{\alpha}
=
\{\alpha_{iJ}\}_{i,J=1}^{3},
\qquad
\alpha_{iJ} \in (0,1),
\end{equation}
and define a total miscoverage budget as the sum over all stress components,
\[
\alpha \triangleq \sum_{i,J} \alpha_{iJ}.
\]
This aggregated quantity serves as a conservative upper bound on the overall miscoverage probability, without
requiring independence among stress components.

The desired coverage requirement is therefore formulated in a marginal sense as
\begin{equation}
\mathbb{P}\!\left(
\boldsymbol{P}^{\ast} \in \mathcal{U}(\boldsymbol{F}^{\ast})
\right)
\ge
1 - \alpha,
\label{eq:cond_cov_tensor}
\end{equation}
where the probability is taken with respect to the joint distribution $p_{\mathrm{data}}(\boldsymbol{F},\boldsymbol{P})$.

The set $\mathcal{U}(\boldsymbol{F})$ is thus a \emph{marginal predictive interval}: it controls uncertainty averaged
over the joint distribution of deformations and stresses, rather than guaranteeing coverage conditional on a fixed
deformation gradient.
Our goal is to construct such marginal predictive intervals in a \emph{distribution-free} manner, in the sense of
conformal prediction, requiring no parametric assumptions on $p_{\mathrm{data}}$ beyond exchangeability.

\begin{figure}[h]
  \centering
  \begin{subfigure}[b]{0.6\textwidth}
    \includegraphics[width=\textwidth]{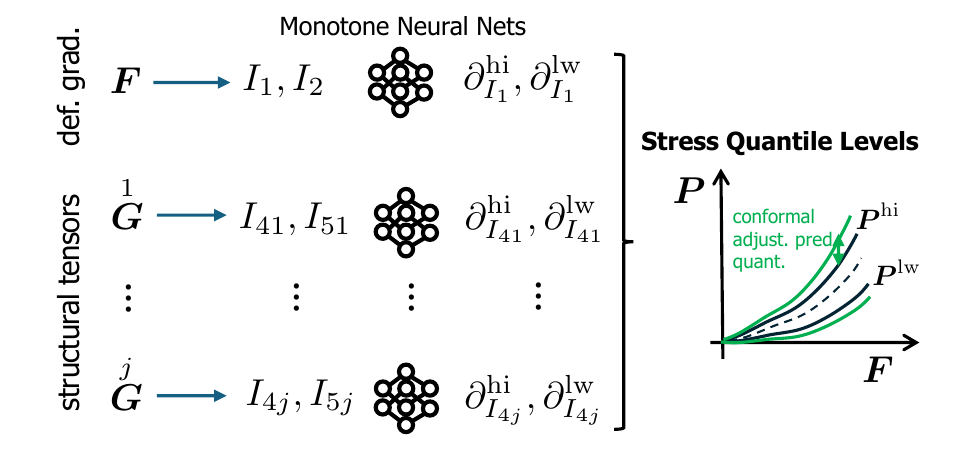}
    %\caption{}
    %\label{subfig:one}
  \end{subfigure}
    \caption{
    \alert{Schematic of the proposed probabilistic physics-constrained neural constitutive modeling framework.
    Invariant features ($I_k$) derived from deformation gradient $\boldsymbol{F}$ and structural tensors $\overset{j}{\boldsymbol{G}}$ are passed through monotone neural networks to learn bounds on strain-energy derivatives. These bounds induce stress predictions with quantile estimates, which are calibrated using conformal prediction for reliable uncertainty quantification.
    }
    }
  \label{fig:method-overview}
\end{figure}

To this end, we adopt a tensorial extension of quantile regression in which componentwise conditional stress quantiles
are estimated using a structured, physics-consistent constitutive representation. \alert{A schematic representation of the overall proposed framework is shown in \cref{fig:method-overview}.}

{\color{Rtwo} 
\remark{
A brief summary of the probabilistic terminologies used frequently in this work is provided in Appendix~\ref{app:prob-terms}.
}
}

\subsubsection{Tensorial Quantile Functions}

%Because the stress response is tensor-valued, uncertainty quantification must ultimately produce predictions in$\mathbb{R}^{3\times 3}$.
Let $P_{iJ}$ denote the $(i,J)$ entry of the stress tensor $\boldsymbol{P}$.
To formally describe probabilistic ordering in this tensor-valued setting, we introduce the tensorial cumulative
distribution function (CDF)
\begin{equation}
\mathrm{CDF}(\boldsymbol{P} \mid \boldsymbol{F})
\triangleq
\mathbb{P} (
\bigcap_{i,J}
\{\bar{P}_{iJ} \le P_{iJ}\}
|
\boldsymbol{F}
),
\label{eq:tensorial_cdf}
\end{equation}
which represents the probability that all stress components lie below their specified values. \textcolor{Rone}{Here, $\bigcap$ denotes the intersection of events.} 
This definition induces a partial order on $\mathbb{R}^{3\times 3}$ and serves as a conceptual reference for defining
multivariate quantiles, rather than an object to be estimated directly.

Given a total miscoverage budget $\alpha$ associated with the tensor of componentwise miscoverage levels
$\boldsymbol{\alpha}$, a corresponding tensorial quantile surface can be formally defined as
\begin{equation}
\boldsymbol{q}_{\boldsymbol{\alpha}}(\boldsymbol{F})
\triangleq
\inf\left\{
\boldsymbol{P} \in \mathbb{R}^{3\times 3}
:\;
\mathrm{CDF}(\boldsymbol{P} \mid \boldsymbol{F})
\ge
1-\alpha
\right\},
\label{eq:tensorial_quantile}
\end{equation}
which defines a lower boundary in the nine-dimensional stress space with respect to the induced partial order.
In general, this tensorial quantile surface depends on the full joint conditional distribution of the stress components
and is neither unique nor directly accessible in practice.

Rather than attempting to estimate such a joint multivariate quantile, we adopt a conservative and practically
tractable componentwise construction.
Specifically, we split each component’s miscoverage \textit{symmetrically} as
\begin{equation}
\alpha_{iJ}^{\mathrm{lo}} = \frac{1}{2}\alpha_{iJ},
\qquad
\alpha_{iJ}^{\mathrm{hi}} = 1 - \frac{1}{2}\alpha_{iJ},
\label{eq:tensorial_alpha_lo_hi}
\end{equation}
which defines lower and upper quantile levels for each stress component.

This yields a componentwise lower quantile surface
$\boldsymbol{q}_{\boldsymbol{\alpha}^{\mathrm{lo}}}(\boldsymbol{F})$
and an upper quantile surface
$\boldsymbol{q}_{\boldsymbol{\alpha}^{\mathrm{hi}}}(\boldsymbol{F})$.
The resulting predictive set is defined as the Cartesian product of these componentwise intervals,
\begin{equation}
\mathcal{U}(\boldsymbol{F})
=
\big[
\boldsymbol{q}_{\boldsymbol{\alpha}^{\mathrm{lo}}}(\boldsymbol{F}),
\;
\boldsymbol{q}_{\boldsymbol{\alpha}^{\mathrm{hi}}}(\boldsymbol{F})
\big],
\label{eq:U_interval}
\end{equation}
which constitutes a marginal, distribution-free predictive interval for the tensor-valued stress response.
This construction does not require modeling the joint dependence structure among stress components and enables
finite-sample coverage guarantees through a union-bound argument, see Appendix \ref{appx:proof-qr-covg}.
\textcolor{Rone}{From an engineering standpoint, the resulting quantile surfaces provide conservative, componentwise envelopes that may inform reliability-aware or worst-case design decisions. The width of these intervals also reflects the level of predictive uncertainty under the available data.}

\subsubsection{Quantile Regression for Tensor-Valued Stress Prediction}

To obtain the tensorial quantile surfaces introduced above, we require a parametric model that maps a deformation
gradient $\boldsymbol{F}$ to stress predictions associated with prescribed quantile levels.
Following the additive structure of the deterministic constitutive backbone, we represent the
$\alpha$-level stress quantile using a structured additive parameterization,
\begin{equation}
q_{\alpha}(\boldsymbol{F};\theta)
=
\sum_{k=1}^{M}
\frac{\partial \tilde{I}_{k}}{\partial \boldsymbol{F}}
\,
\tilde{g}_{\alpha k}(\tilde{I}_{k};\theta_{k}),
\label{eq:q_alpha_model}
\end{equation}
where $\tilde{g}_{\alpha k}$ denotes a quantile-indexed family of monotone scalar functions.
While the quantile of a sum is not, in general, equal to the sum of the corresponding quantiles, this additive
representation is adopted as a \emph{physically consistent inductive bias} that preserves the structure of
hyperelasticity, material symmetry, and thermodynamic admissibility.
Within this constrained hypothesis class, the model seeks the best approximation of the desired quantile functional.
%in the sense induced by the pinball loss.

Given training samples $(\boldsymbol{F}^{(n)}, \boldsymbol{P}^{(n)}) \sim \mathcal{D}_{\mathrm{trn}}$, the parameters
$\theta = \{\theta_k\}$ are estimated by minimizing the expected pinball loss~\cite{koenker2001quantile},
\begin{equation}
\theta
=
\arg\min_{\hat{\theta}}
\;
\mathbb{E}_{(\boldsymbol{F},\boldsymbol{P})\sim\mathcal{D}_{\mathrm{trn}}}
\!\left[
\sum_{i,J}
\rho_{\alpha_{iJ}}\!\left(
P_{iJ},
q_{\alpha,iJ}(\boldsymbol{F};\hat{\theta})
\right)
\right],
\label{eq:quantile_regression_obj}
\end{equation}
where the loss is evaluated componentwise.
For any scalar response $y$ and prediction $\hat{y}$, the pinball loss at quantile level $\alpha$ is defined as
\begin{equation}
\rho_{\alpha}(y,\hat{y})
=
\left(\alpha - \mathbb{I}\{y - \hat{y} < 0\}\right)
(y - \hat{y}),
\label{eq:pinball_loss}
\end{equation}
which is differentiable almost everywhere.

This optimization procedure does not attempt to learn the full conditional distribution of the stress response.
Rather, it identifies quantile functions that best match the observed stress data under the chosen structured
constitutive representation.
%As a result, the learned mapping $q_{\alpha}(\boldsymbol{F};\theta)$ should be interpreted as a physics-consistent quantile predictor, rather than an exact representation of the true conditional quantile surface.

For each prescribed quantile level, the invariant-gradient functions
$\{\tilde g_{\alpha k}\}_{k=1}^{M}$ are learned \emph{jointly} using all stress components through the aggregated
pinball loss.
In particular, training at the lower and upper quantile levels
$\boldsymbol{\alpha}^{\mathrm{lo}}$ and $\boldsymbol{\alpha}^{\mathrm{hi}}$ produces two tensor-valued quantile
predictors that share the same constitutive structure and couple all stress components through the underlying
hyperelastic formulation.
These predictors together define the initial tensor-valued predictive set
$\mathcal{U}(\boldsymbol{F})$ prior to conformal calibration.

\paragraph{Quantile Monotonicity.}
To ensure the logical consistency of the predictive intervals, we must prevent \emph{quantile crossing}, whereby the
predicted lower quantile $\hat{q}_{\boldsymbol{\alpha}^{\mathrm{lo}}}$ exceeds the upper quantile
$\hat{q}_{\boldsymbol{\alpha}^{\mathrm{hi}}}$.
In addition to the pinball loss, we therefore introduce a non-crossing penalty,
\begin{equation}
\mathcal{L}_{\mathrm{nc}}
=
\frac{1}{N}
\sum_{n=1}^{N}
\max\!\left(
0,
\hat{q}_{\boldsymbol{\alpha}^{\mathrm{lo}}}(\boldsymbol{F}^{(n)})
-
\hat{q}_{\boldsymbol{\alpha}^{\mathrm{hi}}}(\boldsymbol{F}^{(n)})
\right).
\end{equation}
This penalty acts as a soft constraint during training and enforces a valid ordering of the predicted quantile
functions across the deformation domain.
In all numerical experiments presented in this work, this term remained identically zero,
indicating that the monotone invariant-based parameterization may naturally preserves the ordering of the quantile
predictors.

\subsubsection{Conformal Adjustment for Distribution-Free Coverage}
\label{sec:conformal}

The quantile-regression model described above provides estimates of lower and upper stress quantiles at prescribed
levels through a \emph{single, jointly trained tensor-valued constitutive model}.
However, in finite-data settings, these raw quantile predictions do not in general achieve the desired marginal
coverage $1-\alpha_{iJ}$ for each stress component due to model misspecification, limited sample size, or violations
of idealized population assumptions.

To restore finite-sample, distribution-free coverage guarantees at the component level, we apply a
\emph{componentwise} version of conformalized quantile regression (CQR)~\cite{romano2019conformalized}.
Importantly, this conformal adjustment is applied \emph{after} training and does not alter the underlying
physics-consistent quantile models.
Instead, it calibrates the predicted quantile intervals using a held-out calibration set
$(\boldsymbol{F}^{(n)},\boldsymbol{P}^{(n)}) \in \mathcal{D}_{\mathrm{cal}}$
and yields marginal coverage guarantees for each stress component under the assumption of exchangeability.

For each component $(i,J)$ and each calibration sample, we define the nonconformity score
\begin{equation}
e^{(n)}_{iJ}
\triangleq
\max\!\big(
\hat{q}_{\alpha^{\mathrm{lo}}_{iJ}}(\boldsymbol{F}^{(n)}) - P^{(n)}_{iJ},
\;
P^{(n)}_{iJ} - \hat{q}_{\alpha^{\mathrm{hi}}_{iJ}}(\boldsymbol{F}^{(n)}),
\;
0
\big),
\label{eq:conformal_scores_componentwise}
\end{equation}
which measures the extent to which the observed stress component falls outside the predicted quantile interval.
If the observation lies within the interval, the score is zero.

The empirical conformal quantile for component $(i,J)$ is then computed as
\begin{equation}
Q_{1-\alpha_{iJ}}(e_{iJ},\mathcal{D}_{\mathrm{cal}})
\triangleq
\mathrm{Quantile}_{(1-\alpha_{iJ})(1+1/|\mathcal{D}_{\mathrm{cal}}|)}
\big(
e^{(1)}_{iJ},\dots,e^{(|\mathcal{D}_{\mathrm{cal}}|)}_{iJ}
\big).
\end{equation}

For a new deformation gradient $\boldsymbol{F}$, the calibrated predictive interval for each stress component is
defined as
\begin{equation}
C_{iJ}(\boldsymbol{F})
\triangleq
\big[
\hat{q}_{\alpha^{\mathrm{lo}}_{iJ}}(\boldsymbol{F}) - Q_{1-\alpha_{iJ}},
\;
\hat{q}_{\alpha^{\mathrm{hi}}_{iJ}}(\boldsymbol{F}) + Q_{1-\alpha_{iJ}}
\big].
\end{equation}
The full tensor-valued predictive set is obtained as the Cartesian product of the componentwise intervals,
\begin{equation}
C(\boldsymbol{F})
\triangleq
\left\{
\boldsymbol{P}\in\mathbb{R}^{3\times 3}
:
P_{iJ}\in C_{iJ}(\boldsymbol{F})
\;\;\text{for all}\;\; i,J
\right\},
\label{eq:componentwise_conformal_tensor_set}
\end{equation}
which provides a conservative uncertainty set for the tensor-valued stress response without modeling the joint
dependence structure among components.

%\paragraph{Finite-sample marginal coverage guarantee.}
By the main theorem of~\cite{romano2019conformalized}, if the calibration data are exchangeable, then for each stress
component $(i,J)$,
\[
\mathbb{P}\!\left( P_{iJ} \in C_{iJ}(\boldsymbol{F}) \right) \ge 1 - \alpha_{iJ},
\]
where the probability is taken with respect to the joint distribution of
$(\boldsymbol{F},\boldsymbol{P})$.
Consequently, the tensor-valued predictive set $C(\boldsymbol{F})$, defined as the Cartesian product of the
componentwise conformal intervals, achieves finite-sample, distribution-free, componentwise marginal coverage; see Appendix \ref{appx:proof-cqr} for the proof sketch.

\textcolor{Rtwo}{
\remark{
Our quantile and conformal calibration procedures are applied to each scalar stress component $P_{iJ}$ \emph{conditional on the full deformation gradient} $\boldsymbol{F}$ (i.e., $\boldsymbol{F}$ is treated as the covariate and no independence between its entries is assumed). Moreover, the resulting tensor-valued predictive set is constructed as a Cartesian product of componentwise intervals; this is a conservative construction whose overall miscoverage is controlled via a union-bound argument and therefore does \emph{not} require any independence assumptions among stress components (see Appendix~\ref{appx:proof-qr-covg}).}}

\section{Numerical Examples}
\label{sec:examples}
In this section, three numerical examples are presented to demonstrate and benchmark the performance of the proposed method. In the first problem, we synthesize data from a classical constitutive model with arbitrarily chosen parameters. In the second problem, we generate synthetic data using a recently introduced constitutive model with parameters calibrated based on experimental measurements. Finally, in the third problem, we apply the proposed method directly to experimental data reported in the literature. The details of each problem are provided in the corresponding subsections, and additional information on training procedures and model parameters is given in Appendix~\ref{appx:params-all}. For completeness, the different experimental loading conditions considered in this section are summarized in Appendix~\ref{appx:loading_conditions}.

\textbf{A Note on Experimental Loading Conditions}

Although the hyperelastic constitutive model considered in this work is path-independent, experimental stress–strain data are typically collected along prescribed loading trajectories. As a result, even in the absence of intrinsic path dependence in the constitutive law, the observed stress states within a single loading trajectory are strongly correlated due to the experimental protocol and sequential nature of data acquisition.

The conformal prediction framework described in the previous section provides distribution-free marginal coverage guarantees under the assumption of exchangeability. 
{\color{Rone}
A natural interpretation of this assumption is a \emph{pooled (pointwise) calibration} strategy, in which all observations across trajectories are treated as exchangeable samples and nonconformity scores are computed at the level of individual loading steps.}
However, this assumption may be inappropriate in the presence of strong intra-trajectory correlations. {\color{Rone}When data are collected along structured, prescribed loading paths (e.g., a monotonically increasing uniaxial test), treating individual stress measurements as exchangeable samples can lead to overly optimistic (i.e., under-covered) uncertainty estimates.} In such settings, a more appropriate exchangeability assumption is at the level of entire loading trajectories.

Accordingly, in the numerical examples presented in this work, we adopt a \emph{trajectory-level conformal calibration} strategy {\color{Rone} due to the use of data generated from structured loading paths.} Specifically, for each calibration trajectory
\[
\mathcal{T}^{(m)} = \{(\boldsymbol{F}^{(m)}(t), \boldsymbol{P}^{(m)}(t))\}_{t=1}^{T_m},
\]
we compute pointwise nonconformity scores
\[
e^{(m)}(t)
=
\max\!\left(
\hat{q}^{\mathrm{lo}}(\boldsymbol{F}^{(m)}(t)) - \boldsymbol{P}^{(m)}(t),
\;
\boldsymbol{P}^{(m)}(t) - \hat{q}^{\mathrm{hi}}(\boldsymbol{F}^{(m)}(t)),
\;
0
\right),
\]
and aggregate them into a single trajectory-level score via
\begin{equation}
e_{\mathrm{traj}}^{(m)}
=
\max_{t \in \mathcal{T}^{(m)}} e^{(m)}(t).
\label{eq:traj-conf}
\end{equation}

{\color{Rone}
This trajectory-level formulation applies the conformal adjustment defined in Section~\ref{sec:conformal} to the aggregated scores $\{e_{\mathrm{traj}}^{(m)}\}$, rather than to individual pointwise scores.}
Conformal calibration is then performed across trajectories by computing the conformal quantile of the set $\{e_{\mathrm{traj}}^{(m)}\}$ over the calibration trajectories, under the assumption that the trajectories themselves are exchangeable.

This procedure yields a calibrated uncertainty set such that, with probability at least $1-\alpha$, the predicted stress interval contains the entire stress response along a previously unseen loading trajectory. Importantly, this trajectory-level calibration does not alter the underlying hyperelastic model or the quantile regression formulation introduced in Section~\ref{sec:conformal}; rather, it provides a practically meaningful calibration and evaluation protocol that accounts for correlation induced by trajectory-wise data collection.

\subsection{Synthetic Data of the Mooney–Rivlin Model}
In this example, we study the proposed method in an isotropic setting to evaluate the fully in-distribution performance of the probabilistic framework.
The corresponding statistical setting represents a pure interpolation regime, in which all realizations are assumed to be sampled from the same underlying probability measure and the loading paths remain within the same range across all realizations.

\begin{figure}[h]
  \centering
  \begin{subfigure}[b]{0.3\textwidth}
    \includegraphics[width=\textwidth]{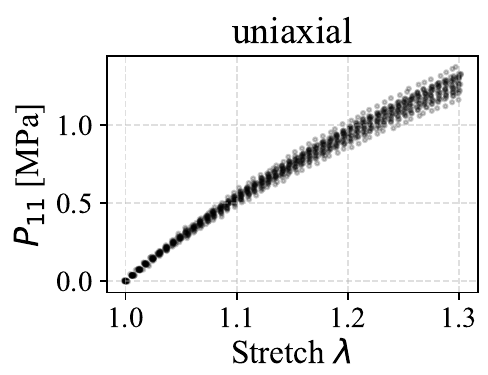}
    %\caption{}
    %\label{subfig:one}
  \end{subfigure}
    \begin{subfigure}[b]{0.3\textwidth}
    \includegraphics[width=\textwidth]{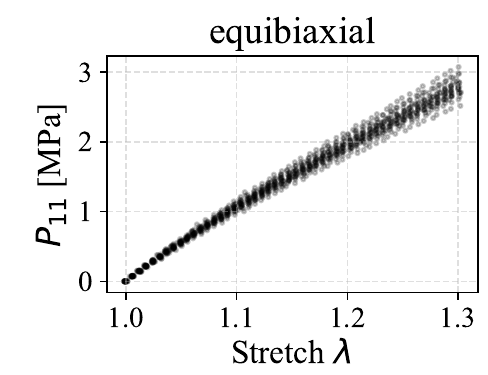}
    %\caption{}
    %\label{subfig:one}
  \end{subfigure}
 \begin{subfigure}[b]{0.3\textwidth}
    \includegraphics[width=\textwidth]{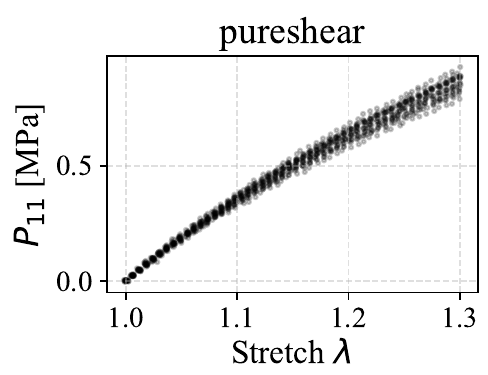}
    %\caption{}
    %\label{subfig:one}
  \end{subfigure}
  \begin{subfigure}[b]{0.3\textwidth}
    \includegraphics[width=\textwidth]{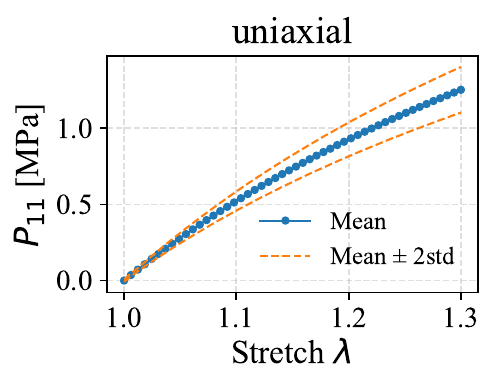}
    %\caption{}
    %\label{subfig:one}
  \end{subfigure}
    \begin{subfigure}[b]{0.3\textwidth}
    \includegraphics[width=\textwidth]{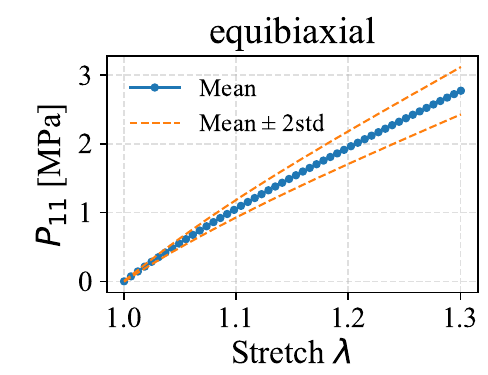}
    %\caption{}
    %\label{subfig:one}
  \end{subfigure}
 \begin{subfigure}[b]{0.3\textwidth}
    \includegraphics[width=\textwidth]{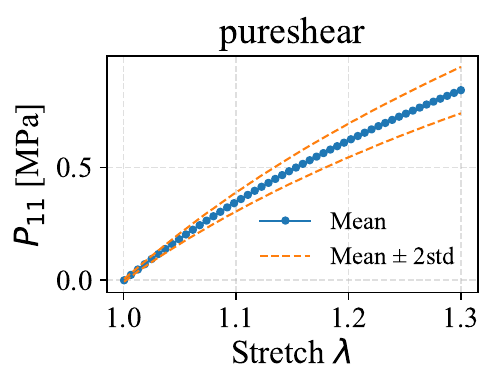}
    %\caption{}
    %\label{subfig:one}
  \end{subfigure}
    \caption{
    Synthetic datasets generated using the Mooney--Rivlin constitutive model. (top) Twenty sets of material parameters are randomly sampled{\color{Rtwo}, with black dots indicating the corresponding data points.} (bottom) Mean and standard deviation computed over 500 parameter realizations.
    }
  \label{fig:MR-data}
\end{figure}

The data are synthesized using the isotropic Mooney--Rivlin constitutive model, with strain--energy density
\begin{equation}
\psi = C_1 \left( I_1 - 3 \right) + C_2 \left( I_2 - 3 \right),
\end{equation}
where $C_1$ and $C_2$ denote material parameters. The data-generating process assumes that these parameters are independent normally distributed random variables with mean $1.0~\mathrm{MPa}$ and standard deviation $0.08~\mathrm{MPa}$. Representative samples of the generated data, together with Monte Carlo estimates of the mean and standard deviation, are shown in Figure~\ref{fig:MR-data}.

Although the problem setup assumes a homogeneous random process in the model-parameter space, the resulting stress–strain responses are heteroscedastic due to the nonlinearity of the constitutive model, as evidenced by the increase in standard deviation with increasing load observed in the figure.

\subsubsection{Conformal Calibration Strategy}
To study the impact of different conformal calibration choices discussed in the previous section, and to justify the trajectory-based approach (see Equation~\ref{eq:traj-conf}) over the pooled approach (see Equation~\ref{eq:traj-conf}) for experimentally sampled data, we conduct a series of numerical experiments on uniaxial data.

\begin{figure}[h]
  \centering
  \begin{subfigure}[b]{0.4\textwidth}
    \includegraphics[width=\textwidth]{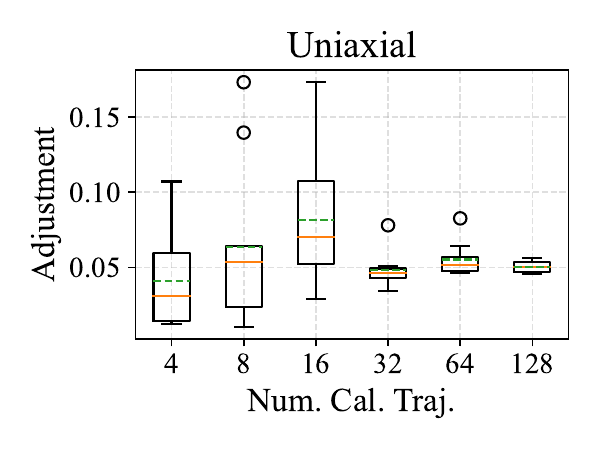}
    %\caption{}
    %\label{subfig:one}
  \end{subfigure}
    \begin{subfigure}[b]{0.4\textwidth}
    \includegraphics[width=\textwidth]{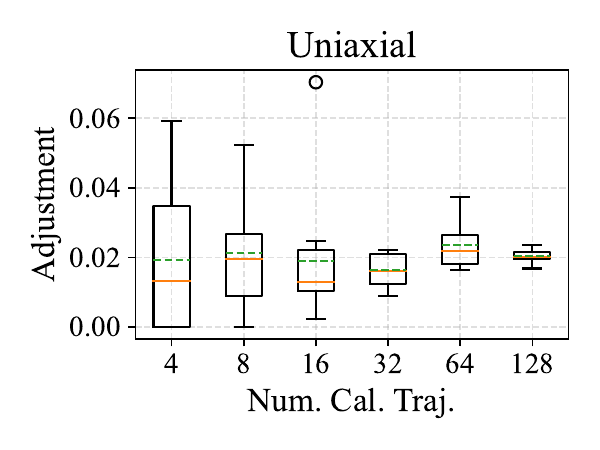}
    %\caption{}
    %\label{subfig:one}
  \end{subfigure}
    \caption{
    Conformal adjustments computed using two calibration strategies: (a) trajectory-wise calibration and (b) pooled calibration across all trajectories. The horizontal axis denotes the number of calibration trajectories randomly selected from a batch of 256 trajectories. For each setting, the statistics of the conformal adjustments are computed over 10 independent random subsamples drawn from the calibration batch.
    }
  \label{fig:traj-vs-pool-uniax-MR}
\end{figure}

A summary of this comparison is shown in Figure~\ref{fig:traj-vs-pool-uniax-MR}. A single quantile model is trained using a dataset consisting of 15 randomly sampled trajectories. Conformal adjustments are then computed using calibration sets containing varying numbers of trajectories, randomly drawn from a batch of 512 datasets not used during training or testing. As expected, the trajectory-aggregated calibration yields more conservative adjustments, reflected by larger conformal correction values. Consequently, for the remainder of this paper, we adopt this calibration strategy, as all datasets are assumed to be collected under common experimental loading scenarios.

\subsubsection{Conformalized Quantile Prediction under Mixed Loading Conditions}
\begin{figure}[h]
  \centering
  \begin{subfigure}[b]{0.5\textwidth}
    \includegraphics[width=\textwidth]{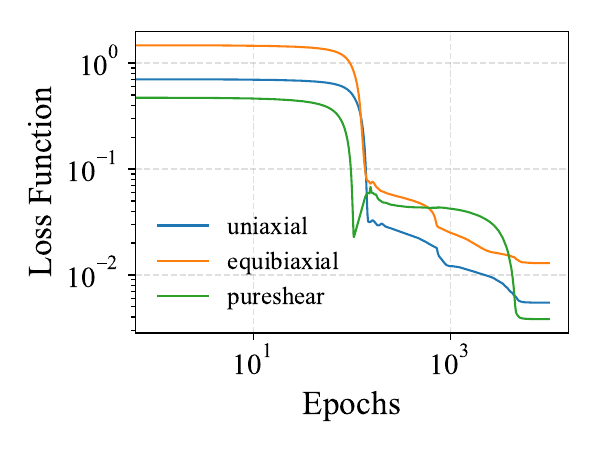}
    %\caption{}
    %\label{subfig:one}
  \end{subfigure}
    \caption{
    Pinball loss for each loading case during training.
    }
  \label{fig:MR-QR-loss}
\end{figure}

We compare the quantile and conformalized quantile predictions for the proposed probabilistic model trained using all loading cases simultaneously. For this purpose, 15 loading trajectories are used for each loading condition as training data. The calibration set had 8 randomly selected loading trajectorirs. The other hyperparameter settings are reported in Appendix \ref{appx:pr-MR}. The evolution of the loss function during training is shown in Figure~\ref{fig:MR-QR-loss}.

\begin{figure}[h]
  \centering
  \begin{subfigure}[b]{0.3\textwidth}
    \includegraphics[width=\textwidth]{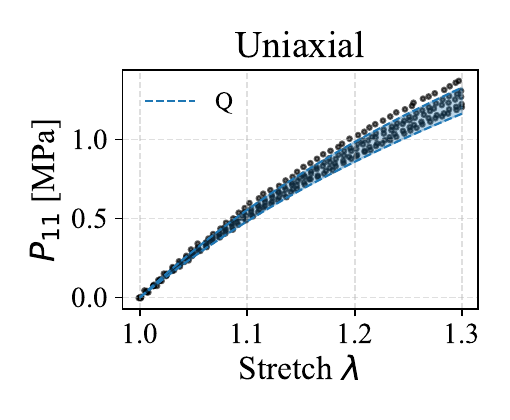}
    %\caption{}
    %\label{subfig:one}
  \end{subfigure}
    \begin{subfigure}[b]{0.3\textwidth}
    \includegraphics[width=\textwidth]{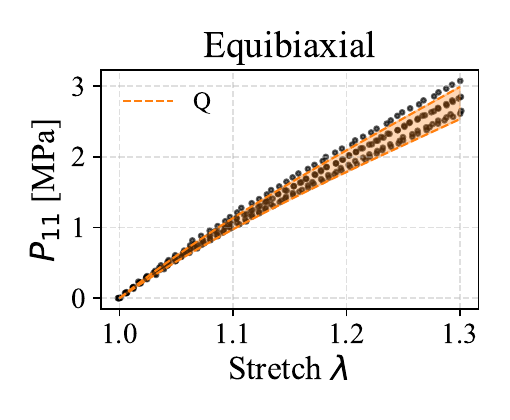}
    %\caption{}
    %\label{subfig:one}
  \end{subfigure}
 \begin{subfigure}[b]{0.3\textwidth}
    \includegraphics[width=\textwidth]{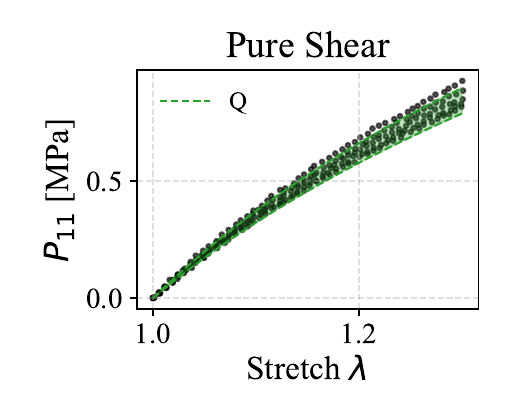}
    %\caption{}
    %\label{subfig:one}
  \end{subfigure}
  \centering
  \begin{subfigure}[b]{0.3\textwidth}
    \includegraphics[width=\textwidth]{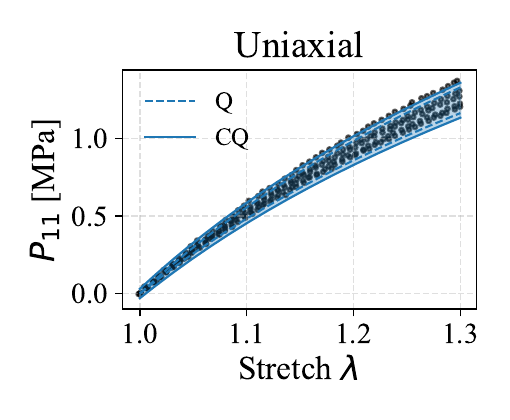}
    %\caption{}
    %\label{subfig:one}
  \end{subfigure}
    \begin{subfigure}[b]{0.3\textwidth}
    \includegraphics[width=\textwidth]{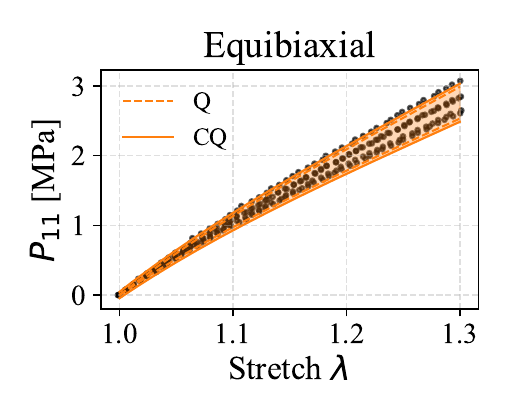}
    %\caption{}
    %\label{subfig:one}
  \end{subfigure}
 \begin{subfigure}[b]{0.3\textwidth}
    \includegraphics[width=\textwidth]{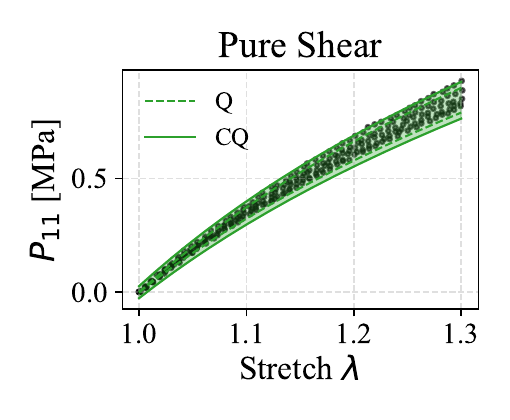}
    %\caption{}
    %\label{subfig:one}
  \end{subfigure}
    \caption{
    (top) Quantile regression uncertainty intervals. (bottom) Conformalized quantile regression. Dashed lines denote the quantile predictions, while solid lines indicate the conformal-adjusted intervals. {\color{Rtwo} Data points are shown as black dots.}
    }
  \label{fig:MR-QR-CQR}
\end{figure}

The quantile predictions, together with their conformal adjustments and the test data (not used in training or calibration), are shown in Figure \ref{fig:MR-QR-CQR}. Quantile regression alone provides a reasonable estimate of predictive intervals that capture both the overall trend and the heteroskedasticity of the data. The conformal adjustment increases the interval width—most noticeably for the uniaxial and pure shear cases—thereby correcting mild underestimation of uncertainty. Nevertheless, a small number of test samples still fall outside the predictive intervals. This behavior is expected in finite-sample settings and reflects the marginal nature of the conformal coverage guarantee. Moreover, conformalized quantile regression accounts only for aleatoric uncertainty; epistemic uncertainty is not modeled in the present work.

\subsection{Synthetic Data of Arterial Wall Material}
In this example, we study the proposed method in an anisotropic setting to evaluate the out-of-distribution performance of the probabilistic framework. The corresponding statistical setting assumes that all realizations are sampled from the same underlying probability measure, while the test loading paths extend beyond the range observed during training, thereby probing the extrapolation capability of the model.

In this problem, we synthesize data using a recently developed constitutive model by Holzapfel et al.~\cite{holzapfel2015modelling}, derived from first principles to capture the complex mechanical behavior of the human arterial wall. This tissue is a soft, fibrous, and anisotropic biological material reinforced by collagen fibre networks.

\begin{figure}[h]
  \centering
  \begin{subfigure}[b]{0.4\textwidth}
    \includegraphics[width=\textwidth]{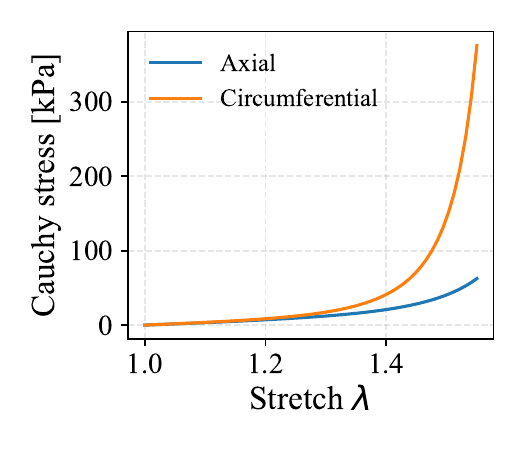}
    %\caption{}
    %\label{subfig:one}
  \end{subfigure}
    \caption{
    Anisotropic material model response with parameters adopted from~\cite{holzapfel2015modelling}, calibrated using arterial wall experimental data.
    }
  \label{fig:data-hoz}
\end{figure}

Since experimental data were not available to us, we employ the model proposed by Holzapfel et al. with the calibrated parameter values reported in their study, which were shown to exhibit excellent agreement with experimental observations (see Figure \ref{fig:data-hoz}). These calibrated models are used to generate synthetic datasets for training and evaluating our framework. %Our objective is to assess whether the proposed method can recover the underlying constitutive behavior in a fully data-driven manner

In the reference model, the strain--energy density function is given by
\begin{equation}
\label{eq:holzapfel_energy}
\Psi
=
\frac{c}{2}\left(\bar I_1 - 3\right)
+
\frac{k_1}{2k_2}
\sum_{i=4,6}
\left\{
\exp\!\left(k_2 \bar E_i^{\,2}\right) - 1
\right\},
\end{equation}
where $c>0$ denotes the shear modulus of the ground matrix, $k_1>0$ is a parameter with dimensions of stress, and $k_2>0$ is a dimensionless parameter. The quantities $\bar E_i$ are Green--Lagrange strain--like measures defined as
\begin{equation}
\label{eq:Ei_invariants}
\bar E_i
=
A\,\bar I_1
+
B\,\bar I_i
+
(1 - 3A - B)\,\bar I_n
-
1,
\end{equation}
with the associated invariants
\begin{equation}
\bar I_i = \bar{\boldsymbol{C}} : (\boldsymbol{M}_i \otimes \boldsymbol{M}_i),
\qquad
\bar I_n = \bar{\boldsymbol{C}} : (\boldsymbol{M}_n \otimes \boldsymbol{M}_n),
\qquad i=4,6.
\end{equation}
Here, $\bar{\boldsymbol{C}} = J^{-2/3}\boldsymbol{C}$ denotes the isochoric part of the right Cauchy--Green tensor. In the reference configuration, the two collagen fiber families are assumed to lie in the circumferential--axial plane and to be symmetrically oriented with respect to the circumferential direction. Their mean directions are given by
\begin{equation}
\boldsymbol{M}_4 = [\cos\alpha,\ \sin\alpha,\ 0]^\mathsf{T},
\qquad
\boldsymbol{M}_6 = [\cos\alpha,\ -\sin\alpha,\ 0]^\mathsf{T},
\end{equation}
while the out-of-plane direction is defined as
\begin{equation}
\boldsymbol{M}_n = [0,\ 0,\ 1]^\mathsf{T}.
\end{equation}
The constants $A$ and $B$ are given by
\begin{equation}
A = 2\,\kappa_{\mathrm{op}}\,\kappa_{\mathrm{ip}},
\qquad
B = 2\,\kappa_{\mathrm{op}}\left(1 - 2\,\kappa_{\mathrm{ip}}\right),
\end{equation}
where $\kappa_{\mathrm{ip}}$ and $\kappa_{\mathrm{op}}$ characterize the in-plane and out-of-plane dispersion of collagen fibers, and $\alpha$ denotes the mean fiber orientation angle. The exponential fiber contributions in~\eqref{eq:holzapfel_energy} are activated only when the corresponding fiber stretches exceed unity ($I_4 > 1$ or $I_6 > 1$), ensuring that collagen fibers contribute to the mechanical response exclusively in tension.

In the synthetic data-generation experiments, we use the calibrated parameter set as the mean of the parameter distribution,
\begin{equation}
\boldsymbol{\mu}
=
(c,\;k_1,\;k_2,\;\kappa_{\mathrm{ip}},\;\kappa_{\mathrm{op}},\;\alpha)^\mathsf{T}
=
(10.07~\text{kPa},\;5.89~\text{kPa},\;21.62,\;0.116,\;0.493,\;47.99^\circ)^\mathsf{T}.
\end{equation}
The stress response of the model with these parameters under uniaxial loading in the circumferential and axial directions is shown in Figure~\ref{fig:data-hoz}.

We artificially introduce noise into the data by assuming that the material parameters are normally distributed around their mean values, with the following standard deviation vector:
\begin{equation}
\boldsymbol{\text{std}}
=
(\text{std}_c,\;\text{std}_{k_1},\;\text{std}_{k_2})^\mathsf{T}
=
(1,\;1,\;1)^\mathsf{T}.
\end{equation}

\begin{figure}[h]
  \centering
  \begin{subfigure}[b]{0.4\textwidth}
    \includegraphics[width=\textwidth]{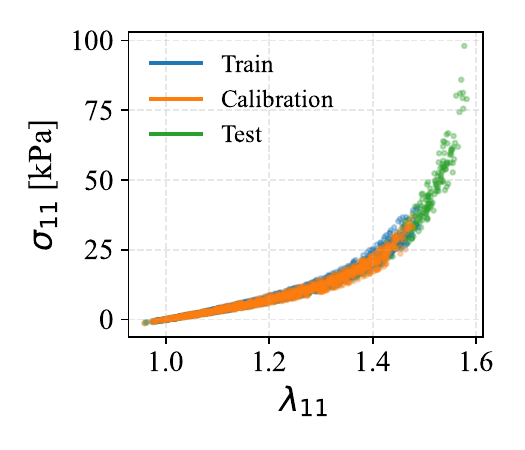}
  \end{subfigure}
  \begin{subfigure}[b]{0.4\textwidth}
    \includegraphics[width=\textwidth]{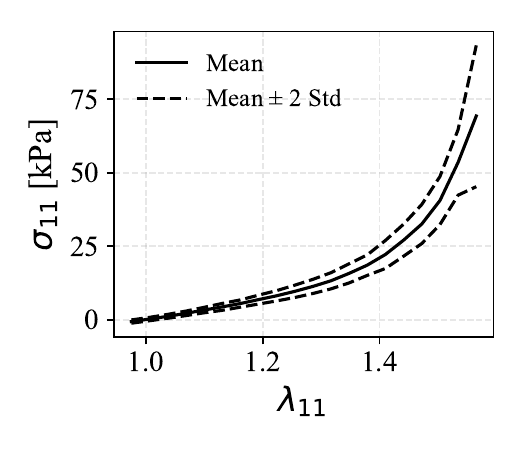}
  \end{subfigure}
  \begin{subfigure}[b]{0.4\textwidth}
    \includegraphics[width=\textwidth]{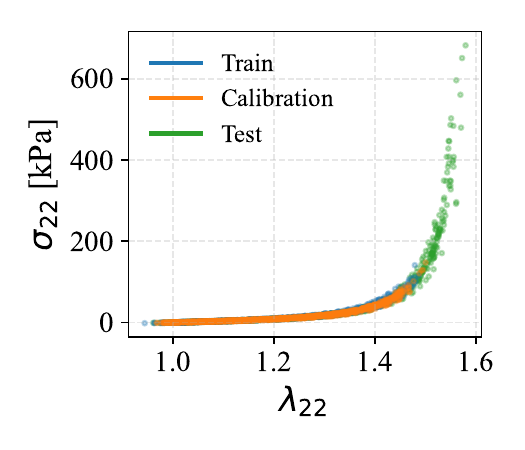}
  \end{subfigure}
  \begin{subfigure}[b]{0.4\textwidth}
    \includegraphics[width=\textwidth]{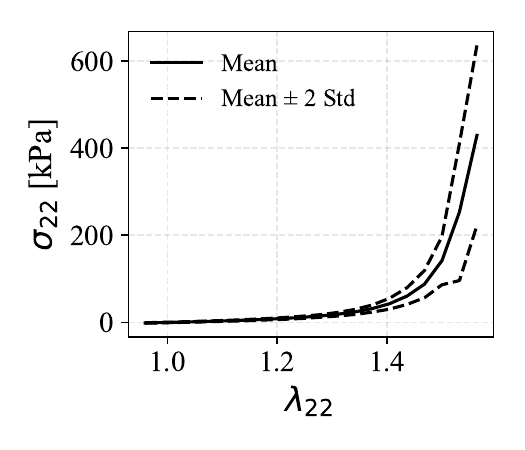}
  \end{subfigure}
    \caption{
    Training, calibration, and test data. (left) Stress–strain responses for uniaxial tests along different directions. (right) Mean and standard deviation of the data as a function of applied strain. {\color{Rtwo} Generated data are shown as colored dots, while solid lines denote the corresponding approximated mean and standard deviation.}
    }
  \label{fig:data-noise-hoz}
\end{figure}

For the training, calibration, and testing datasets, we sample 20 realizations from this distribution to generate uniaxial datasets in the axial and circumferential directions, as shown in Figure~\ref{fig:data-noise-hoz}. Although the data are sampled from a homogeneous random process in the model-parameter space, the resulting stress responses exhibit clear heteroscedasticity \textcolor{Rtwo}{(i.e., non-constant conditional variance)}, as evidenced by the increasing standard deviation with load. This behavior is expected due to the strong nonlinearity of the constitutive model. 
As one of our goals is to evaluate the extrapolation capability of the proposed scheme, the test data are chosen to extend beyond the maximum stretch observed in the training and calibration datasets.

\begin{remark}
The synthesized data are available for download at {\color{Rone}{\texttt{https://doi.org/10.21985/n2-bbg3-zq15}}}.
\end{remark}

\begin{figure}[h]
  \centering
  \begin{subfigure}[b]{0.4\textwidth}
    \includegraphics[width=\textwidth]{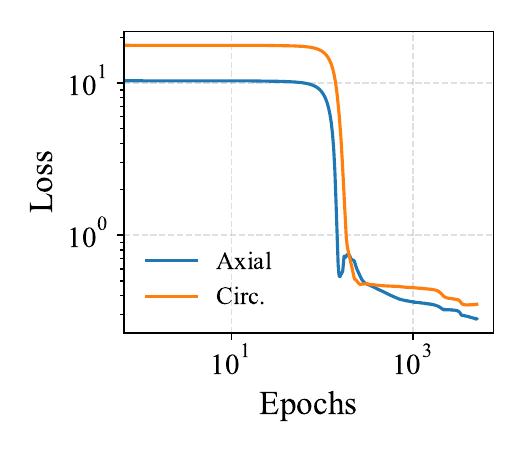}
    %\caption{}
    %\label{subfig:one}
  \end{subfigure}
    \caption{
    Pinball loss training history for the arterial wall data.
    }
  \label{fig:hoz-loss}
\end{figure}

The hyperparameters used in this example are reported in Section~\ref{appx:hoz-data}. During training, both axial and circumferential data are used to learn the probabilistic model, and the training curves of the pinball loss computed for each dataset are shown in Figure~\ref{fig:hoz-loss}.

\begin{figure}[h]
  \centering
  \begin{subfigure}[b]{0.4\textwidth}
    \includegraphics[width=\textwidth]{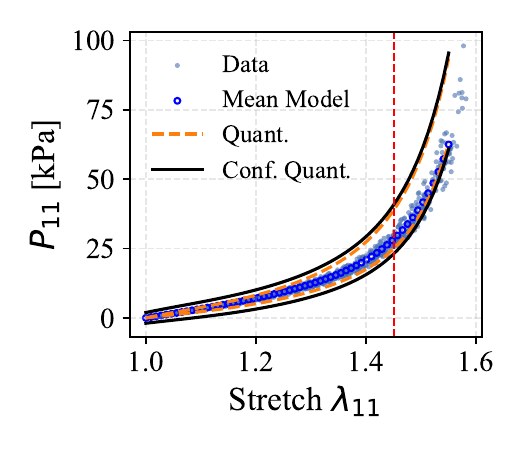}
    %\caption{}
    %\label{subfig:one}
  \end{subfigure}
  \begin{subfigure}[b]{0.4\textwidth}
    \includegraphics[width=\textwidth]{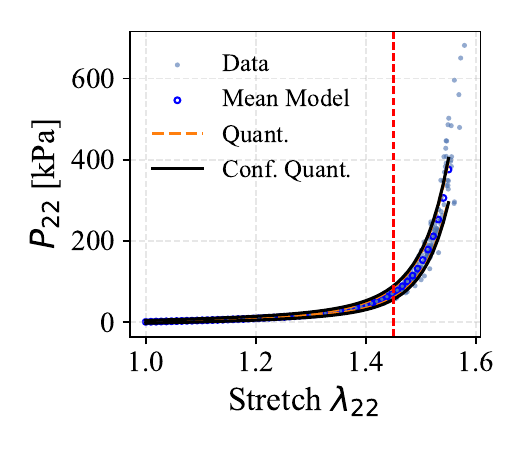}
    %\caption{}
    %\label{subfig:one}
  \end{subfigure}
    \caption{
    Model predictions on the test data (light blue dots) and uncertainty estimates given by the quantile predictions (dashed orange lines) and their conformal adjustments (solid black lines). Blue circles indicate the response corresponding to the mean material parameters of the reference model. The vertical red dashed line denotes the maximum strain used in training.
    }
  \label{fig:hoz-result}
\end{figure}

The response of the trained model on unseen test data extending beyond the maximum strain used for training and calibration is shown in Figure~\ref{fig:hoz-result}. The conformal adjustment contributes more to the $P_{11}$ stress component, whereas the adjustment for $P_{22}$, while positive, is negligible relative to the magnitude of the $P_{22}$ stress. This indicates that, for the given training and calibration datasets, the conformal confidence intervals are already reasonably accurate, with conformalized predictions being only slightly more conservative.

More importantly, the probabilistic model exhibits good extrapolation behavior at higher strain levels, closely following the ground-truth response while providing meaningful uncertainty estimates. At very large strains, however, the estimated uncertainty becomes narrower than the variability implied by the data distribution, indicating the need for additional sampling in these regimes and subsequent retraining of the model. Since conformal guarantees are marginal and depend on the available calibration data, uncertainty estimates in extrapolation regimes motivate adaptive sampling and retraining to improve coverage.

More importantly, the probabilistic model exhibits good extrapolation behavior at higher strain levels, closely following the ground-truth response while providing meaningful uncertainty estimates. At very large strains, however, the estimated uncertainty becomes narrower than the variability implied by the data distribution. 
{\color{Rone}{
This behavior reflects operation in an extrapolation regime beyond the maximum strain used for training and calibration. While additional sampling in these regimes would help improve calibration, structural model discrepancy may also contribute to the observed deviations. Moreover, since the present framework primarily captures aleatoric uncertainty, epistemic effects associated with model inadequacy are not explicitly modeled and may become more pronounced in such extrapolation regimes.
}}
Since conformal guarantees are marginal and depend on the available calibration data, uncertainty estimates in extrapolation regimes motivate adaptive sampling and 
{\color{Rone}{potential model refinement}}
to improve coverage.

\subsection{Experiential Data of Porcine Atrioventricular Valve Leaflets
}
In this example, we demonstrate the ability of the proposed method to infer probabilistic constitutive models for anisotropic biological soft tissues from biaxial experimental data ~\cite{jett2018biaxial}.
%, in settings where no widely accepted classical material model is available in the literature (to the best of our knowledge). 
We consider a loading condition that is not observed during training.

\begin{figure}[h]
  \centering
  \begin{subfigure}[b]{0.9\textwidth}
    \includegraphics[width=\textwidth]{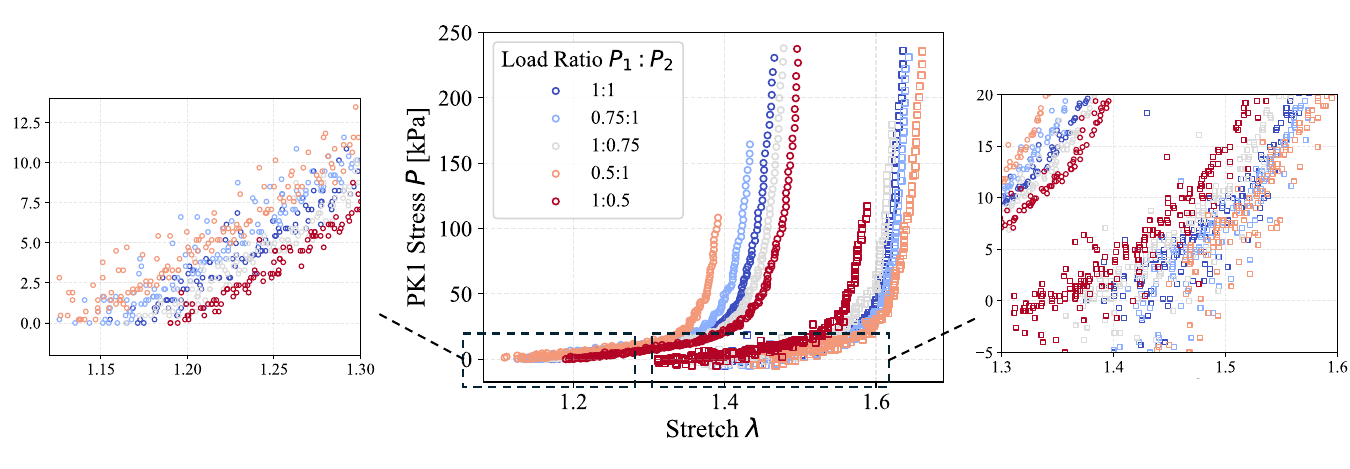}
    %\caption{}
    %\label{subfig:one}
  \end{subfigure}
    \caption{
    Experimental data of porcine atrioventricular valve leaflets under biaxial loading with varying loading ratios, shown in different colors~\cite{jett2018biaxial}. The test data are highlighted in red. {\color{Rtwo}Stress responses along the horizontal direction are denoted by circular markers, while those along the vertical direction are indicated by square markers.}
    }
  \label{fig:real-data}
\end{figure}

The dataset labeled \texttt{MVAL1.txt} from ~\cite{jett2018biaxial} is employed. As shown in \cref{fig:real-data}, the dataset contains axial and circumferential responses under five different biaxial loading ratios. The first four loading conditions are used for training and calibration, while the fifth is reserved for testing. For each loading scenario except the test case, 80\% (randomly selected) of the data are used for training and 20\% for calibration. In this figure, one region from the axial stress response and another from the circumferential response at the onset of loading are highlighted. In these regions, the data exhibit considerable scatter, likely due to measurement noise and the highly compliant nature of the material, which makes it sensitive to small load perturbations. Selected regions of the plot are magnified for clarity.

As mentioned, data corresponding to four different loading conditions are used to train the quantile model. The hyperparameters and training loss histories are reported in Appendix~\ref{appx:porcine-data}.

\begin{figure}[h]
  \centering
  \begin{subfigure}[b]{0.4\textwidth}
    \includegraphics[width=\textwidth]{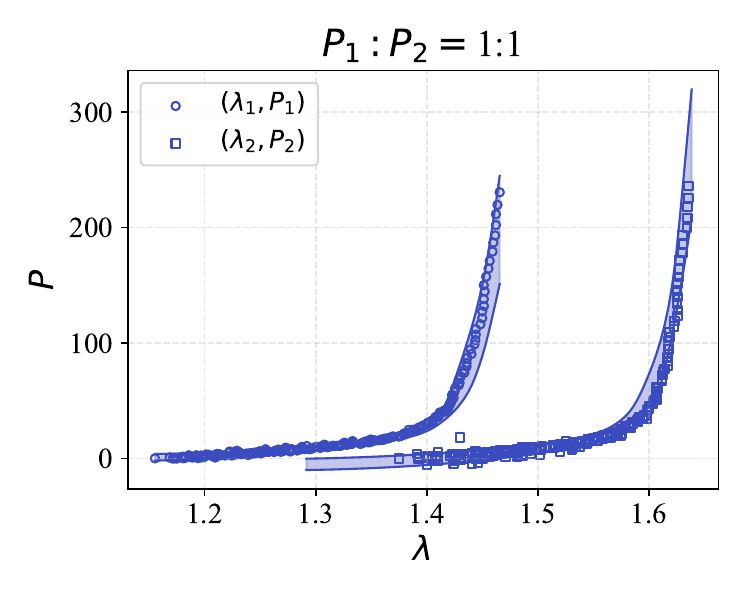}
    %\caption{}
    %\label{subfig:one}
  \end{subfigure}
  \begin{subfigure}[b]{0.4\textwidth}
    \includegraphics[width=\textwidth]{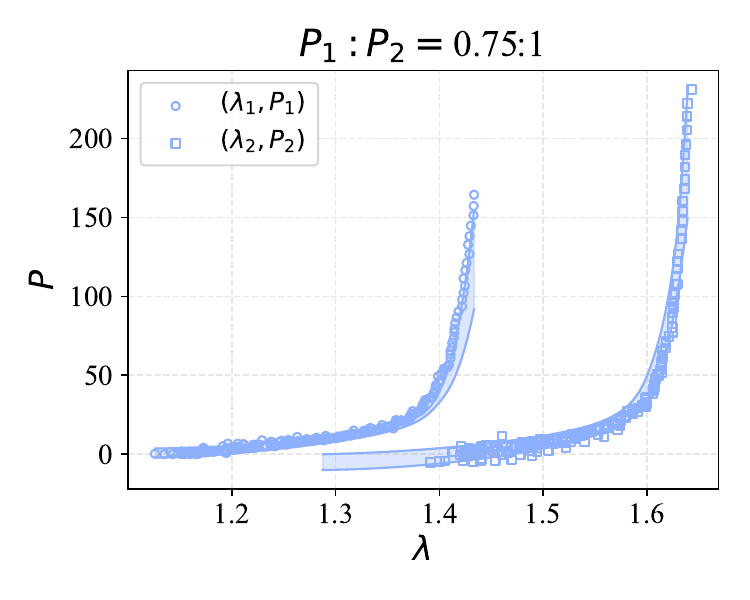}
    %\caption{}
    %\label{subfig:one}
  \end{subfigure}
  \begin{subfigure}[b]{0.4\textwidth}
    \includegraphics[width=\textwidth]{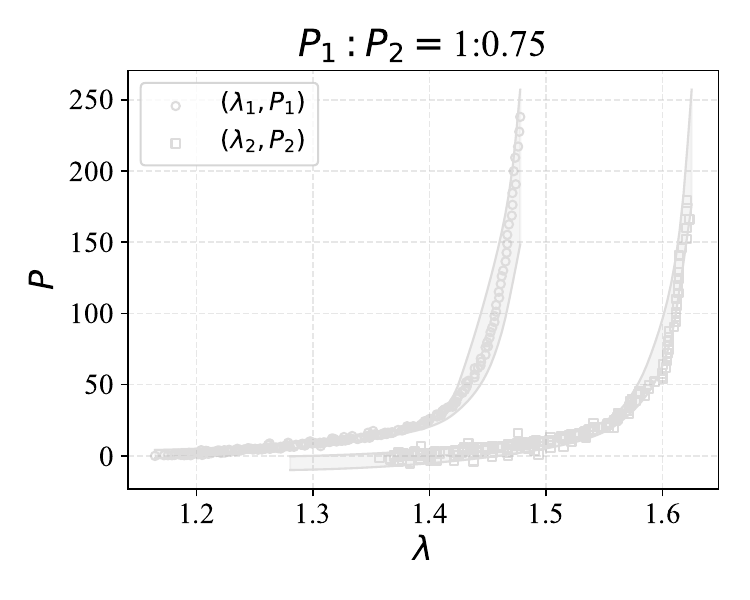}
    %\caption{}
    %\label{subfig:one}
  \end{subfigure}
  \begin{subfigure}[b]{0.4\textwidth}
    \includegraphics[width=\textwidth]{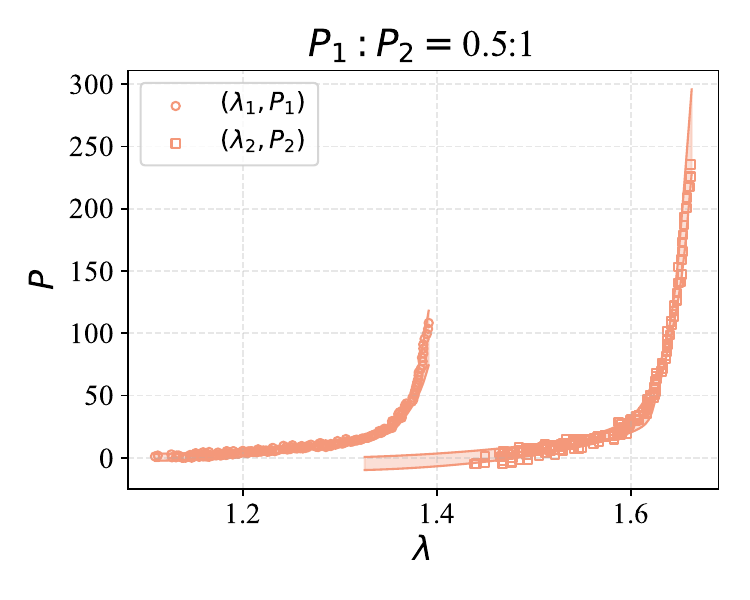}
    %\caption{}
    %\label{subfig:one}
  \end{subfigure}
  \begin{subfigure}[b]{0.4\textwidth}
    \includegraphics[width=\textwidth]{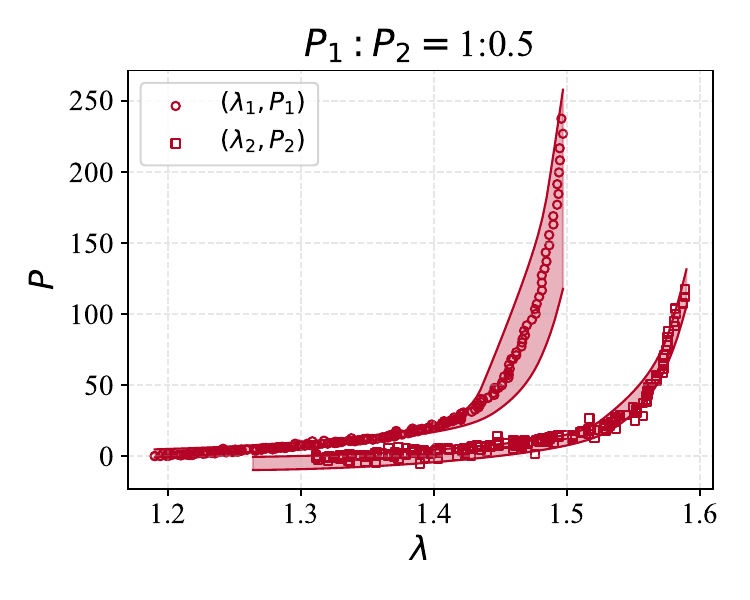}
    %\caption{}
    %\label{subfig:one}
  \end{subfigure}
    \caption{
    {\color{Rtwo}Results of the proposed method after training on experimental data of porcine atrioventricular valve leaflets under biaxial loading. The data are shown using markers, while the solid lines represent the predicted upper and lower quantiles. Circular markers denote stress responses along the horizontal direction, and square markers denote stress responses along the vertical direction. Red indicates results corresponding to unseen test biaxial loading ratios not included during training and calibration.
    }}
  \label{fig:pr3-result}
\end{figure}

Model predictions for each loading condition are shown in Figure~\ref{fig:pr3-result}. The results indicate that the proposed method captures the underlying mechanical response while providing reasonable uncertainty estimates. In certain regions, the predictive intervals appear slightly overconfident, which is attributable to limited data availability; coverage can be improved by enriching the training and calibration datasets.

Importantly, the probabilistic model performs satisfactory on the unseen loading condition (red curves), with uncertainty estimates covering the majority of the experimental data. This behavior is enabled by the physics-encoded parameterization of the anisotropic constitutive response, which enforces consistency with the underlying mechanical and thermodynamic principles. As a result, the learned probabilistic model remains physically admissible while incorporating uncertainty through the learned quantile functions.

\section{Conclusion}
\label{sec:conclud}

In this work, we propose a probabilistic anisotropic constitutive modeling framework that is data-driven while remaining strictly physics-constrained. The proposed probabilistic formulation is frequentist in nature and is based on quantile regression. The unknown quantile functions are parameterized using monotone neural networks that directly represent the quantities required for stress evaluation. This design choice facilitates stable training and ensures that the resulting constitutive model remains thermodynamically consistent across the entire range of probabilistic predictions.

To address potential miscalibration of quantile-based uncertainty estimates arising from finite data, we further introduce a tensor-valued extension of conformalized quantile regression tailored to constitutive modeling applications. This extension enables calibrated uncertainty quantification for stress responses without imposing parametric assumptions on the underlying conditional distributions. An important advantage of the proposed quantile-based formulation is its \textit{plug-and-play} nature: it can be readily integrated into existing deterministic constitutive models, thereby facilitating probabilistic predictive modeling without altering the underlying physics-based formulation.

The proposed framework is benchmarked and demonstrated on several numerical examples, including both synthetic datasets and experimental data, showing predictive performance and uncertainty estimation in both interpolation and extrapolation regimes.

\textbf{Limitations}

While the proposed formulation offers simplicity, ease of training, and fast inference without requiring distributional assumptions, a key limitation is that the resulting conformal prediction intervals primarily capture aleatoric uncertainty associated with data variability. In particular, the approach does not explicitly model epistemic uncertainty arising from model inadequacy or limited data \cite{rossellini2024integrating}. We will leave addressing this issue for {\color{Rone} further} research.

The proposed method, while capable of capturing a wide range of anisotropic behaviors such as transverse isotropy, triclinic, monoclinic, and rhombic symmetries, has also been shown to be insufficient for representing certain other symmetry classes, such as trigonal and hexagonal symmetries. Capturing these additional classes requires extending it to include higher-order structural tensors, such as fourth- and sixth-order tensors~\cite{schroder2008anisotropic}.  
The extension to higher-order formulations is left for future investigation.

{\color{Rone}{
A structural limitation of the present framework is the assumption of additive separability in the invariant-based energy representation (Eq.~\ref{eq:neff}), which excludes explicit cross-invariant coupling effects. Extending the approach to fully coupled multivariate formulations while preserving admissibility guarantees remains an important direction for future work.
}}

%{\color{Rone}{A further structural limitation arises from the assumption of additive separability in the invariant-based energy representation in Equation \ref{eq:neff}. While this structured subclass of polyconvex energy functionals enables monotone univariate parameterizations, tractable enforcement of thermodynamic constraints, and improved data efficiency, it may restrict the ability to represent strong cross-invariant coupling effects. Extending the framework to fully coupled multivariate formulations while preserving admissibility guarantees remains an important direction for future work.}}

%Nevertheless, for the purpose of this work---which aims to introduce a new and efficient probabilistic framework for anisotropic materials---the present second-order formulation is sufficient.

%While we employ separate conformal adjustments for each stress component, the conformal correction within each component is assumed to be constant across the input space, which may not be ideal. Although this assumption simplifies implementation, it can be restrictive in problems exhibiting strong heteroscedasticity, where the magnitude of uncertainty varies significantly with the deformation or loading state. Extensions of conformal prediction that yield heteroscedastic conformal adjustments—often referred to as locally adaptive conformal prediction—have been proposed in the literature \cite{romano2019conformalized}, but are not considered here and are left for future research.

\section*{Data Availability}
All data used in this work are either publicly available or can be reproduced from the information provided and are available from the author upon reasonable request.

\section*{Acknowledgments}

The author gratefully acknowledges the support of Northwestern University through startup research funding.

\section*{Declaration on the Use of Generative AI}

The author used ChatGPT, a large language model developed by OpenAI, to assist with English language editing and grammar refinement. The scientific content, technical interpretations, and conclusions are solely the responsibility of the author.

\appendix

\appendix

{\color{Rtwo}

\section{Supplementary Probabilistic Definitions}
\label{app:prob-terms}

This appendix provides a concise summary of key probabilistic concepts used in the manuscript.

\subsection{Exchangeability}

A sequence $(Z_1, \dots, Z_n)$ is said to be \emph{exchangeable} if
\[
(Z_1, \dots, Z_n) \overset{d}{=} (Z_{\pi(1)}, \dots, Z_{\pi(n)})
\]
for every permutation $\pi$ of $\{1, \dots, n\}$,
where $\overset{d}{=}$ denotes equality in distribution.

In the present setting, we apply this notion to the sequence 
$Z_i = (X_i, Y_i)$ of input–output pairs. Informally, exchangeability means that the samples are statistically indistinguishable and their ordering carries no information. Independent and identically distributed (i.i.d.) samples form a special case of exchangeable sequences.%; however, conformal prediction requires only exchangeability, not independence.
%Exchangeability is weaker than independence but is sufficient to guarantee the finite-sample coverage property of conformal prediction.

\subsection{Nonconformity Scores}

Conformal prediction calibrates predictive intervals using a data-driven
\emph{nonconformity score}, which measures how much an observation
violates a predicted uncertainty set. In the quantile-based setting used in this work, the nonconformity
score for a calibration sample $(x_i, y_i)$ is defined as
\[
e_i
=
\max\big(
\hat{q}_{\alpha^{\mathrm{lo}}}(x_i) - y_i,
\;
y_i - \hat{q}_{\alpha^{\mathrm{hi}}}(x_i),
\;
0
\big),
\]
which quantifies the extent to which the response $y_i$ lies outside
the predicted quantile interval $[\hat{q}_{\alpha^{\mathrm{lo}}}(x_i), \hat{q}_{\alpha^{\mathrm{hi}}}(x_i)]$.
If the observation lies within the interval, the score is zero.

\subsection{Conformal Adjustment}

Let $\{e_i\}_{i=1}^m$ denote the nonconformity scores computed on the calibration set of size $m$. 
The conformal adjustment is defined as
\[
\hat{q}_{\mathrm{adj}}
=
e_{(k)}, 
\qquad
k = \left\lceil (m+1)(1-\alpha) \right\rceil,
\]
where $e_{(k)}$ denotes the $k$-th smallest calibration score.

\subsection{Conformalized Quantile Regression (CQR)}

CQR combines quantile regression with conformal calibration to construct predictive intervals with finite-sample marginal coverage guarantees. Given a trained model that predicts conditional quantiles
\[
\hat{q}_{\alpha^{\mathrm{lo}}}(x), \quad \hat{q}_{\alpha^{\mathrm{hi}}}(x),
\]
nonconformity scores are computed on a held-out calibration set as
\[
e_i = \max\left\{
\hat{q}_{\alpha^{\mathrm{lo}}}(x_i) - y_i,\;
y_i - \hat{q}_{\alpha^{\mathrm{hi}}}(x_i),
\;
0
\right\}.
\]

The conformal adjustment $\hat{q}_{\mathrm{adj}}$ is then defined as the
$\left\lceil (m+1)(1-\alpha) \right\rceil$-th smallest nonconformity score among $\{e_i\}_{i=1}^m$. The final predictive interval is given by
\[
\left[
\hat{q}_{\alpha^{\mathrm{lo}}}(x) - \hat{q}_{\mathrm{adj}}, \quad
\hat{q}_{\alpha^{\mathrm{hi}}}(x) + \hat{q}_{\mathrm{adj}}
\right].
\]

}

\section{Proof of Monotonicity}
\label{app:proof_monotone_net}

We provide a short proof that the univariate network $\mathcal{N}_k:\mathbb{R}\to\mathbb{R}$ defined by
\begin{equation}
h^{(0)} = x,
\qquad
h^{(\ell+1)} = \sigma\!\left(W^{(\ell)} h^{(\ell)} + b^{(\ell)}\right),
\quad \ell=0,\dots,L-1,
\end{equation}
is monotone non-decreasing in $x$ under the assumptions that (i) $W^{(\ell)}$ has elementwise non-negative entries and (ii) the activation $\sigma(\cdot)$ is non-decreasing (equivalently, $\sigma'(z)\ge 0$ wherever the derivative exists).

\paragraph{Claim.}
If $W^{(\ell)} \ge 0$ elementwise for all $\ell$ and $\sigma$ is non-decreasing, then $\mathcal{N}_k(x)=h^{(L)}(x)$ is monotone non-decreasing in $x$, i.e.,
\begin{equation}
x_1 \le x_2 \;\;\Longrightarrow\;\; \mathcal{N}_k(x_1) \le \mathcal{N}_k(x_2).
\end{equation}

\paragraph{Proof.}
We proceed by induction on the layers. Let $x_1\le x_2$ be arbitrary. Since $h^{(0)}(x)=x$, we have $h^{(0)}(x_1)\le h^{(0)}(x_2)$. Assume $h^{(\ell)}(x_1)\le h^{(\ell)}(x_2)$ holds elementwise for some $\ell\in\{0,\dots,L-1\}$. Because $W^{(\ell)}\ge 0$ elementwise, the affine map preserves order:
\begin{equation}
W^{(\ell)} h^{(\ell)}(x_1) + b^{(\ell)}
\;\le\;
W^{(\ell)} h^{(\ell)}(x_2) + b^{(\ell)}, \quad \text{(elementwise.)}
\end{equation}
Since $\sigma$ is non-decreasing, applying $\sigma$ preserves the inequality, yielding
\begin{equation}
h^{(\ell+1)}(x_1)
=
\sigma\!\left(W^{(\ell)} h^{(\ell)}(x_1) + b^{(\ell)}\right)
\;\le\;
\sigma\!\left(W^{(\ell)} h^{(\ell)}(x_2) + b^{(\ell)}\right)
=
h^{(\ell+1)}(x_2), \quad \text{(elementwise.)}
\end{equation}
This completes the induction, and therefore
$h^{(L)}(x_1)\le h^{(L)}(x_2)$, i.e., $\mathcal{N}_k$ is monotone non-decreasing.

\hfill$\square$

\paragraph{Implementation note.}
In the present work, the non-negativity constraint on the weights is enforced by the reparameterization
$W^{(\ell)} = \left(\widehat{W}^{(\ell)}\right)^2$ for unconstrained $\widehat{W}^{(\ell)}$, which guarantees $W^{(\ell)}\ge 0$ by construction and enables unconstrained optimization.

\section{Numerical Illustration of Monotone Neural Network}
\label{appx:monoton}
This appendix summarizes the implementation details of the numerical illustration presented in \cref{fig:mono-vs-convex}, which compares convex-function parameterization and gradient-based monotone parameterization in a univariate setting.

Two neural network models are considered.  
The monotone gradient model (\texttt{Monotone1DNet}) is a univariate feedforward neural network with three hidden layers of width 32, using the sigmoid activation function. 
The convex energy model is implemented as an input convex neural network (ICNN) with one-dimensional input and output, three hidden layers of width 32, and standard ICNN constraints to enforce convexity with respect to the input.

\begin{figure}[h]
  \centering
  \includegraphics[width=0.35\textwidth]{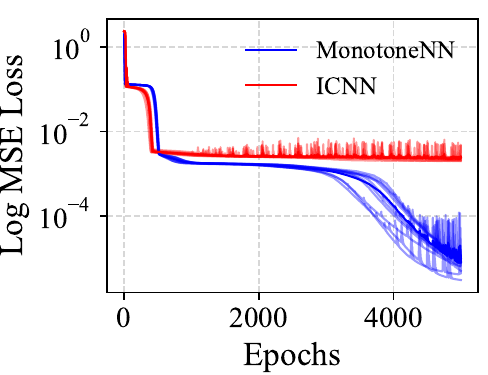}
  \includegraphics[width=0.35\textwidth]{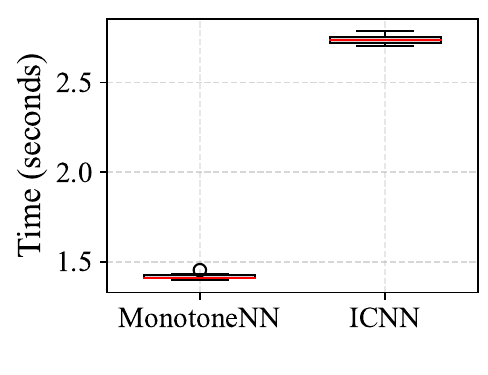}
  \caption{
  Training loss evolution (left) and distribution of training times (right) over 10 randomly initialized neural networks for each parametrization.
  }
  \label{fig:loss-mono-vs-convex}
\end{figure}

Both models are trained to match sampled gradient data using the mean squared error (MSE) loss. Training is performed for 5000 epochs using the Adam optimizer with a learning rate of $10^{-2}$. To account for variability due to random initialization, each experiment is repeated over $10$ independent runs. The training loss evolution and total training time are reported in~\cref{fig:loss-mono-vs-convex}. The monotone parameterization achieves lower training error while requiring approximately half the computational time.

\paragraph{ICNN architecture.}
Let $x \in \mathbb{R}$ denote a scalar strain invariant. The ICNN defines a mapping
\begin{equation}
\tilde{f}_k(x) = \mathcal{F}(x;\theta),
\end{equation}
where $\mathcal{F}:\mathbb{R}\to\mathbb{R}$ is a feedforward neural network composed of affine transformations and convex, non-decreasing activation functions. For a network with $L$ layers, the forward propagation is given by
\begin{equation}
h^{(\ell+1)} = \sigma\!\left( h^{(\ell)} W^{(\ell)}_{+} + b^{(\ell)} \right),
\qquad \ell = 0,\dots,L-2,
\end{equation}
with the final layer defined as
\begin{equation}
\tilde{f}_k(x) = h^{(L-1)} W^{(L-1)}_{+} + b^{(L-1)},
\end{equation}
where $h^{(0)} = x$. The activation function $\sigma(\cdot)$ is chosen to be convex and non-decreasing (e.g., softplus), and the weight matrices satisfy
\begin{equation}
W^{(\ell)}_{+} = \left(\widehat{W}^{(\ell)}\right)^2,
\end{equation}
for unconstrained matrices $\widehat{W}^{(\ell)}$. This construction guarantees elementwise non-negativity of the weights, which, together with the choice of activation function, ensures that $\tilde{f}_k(x)$ is convex with respect to $x$.

\section{Quantile Coverage Guarantee}
\label{appx:proof-qr-covg}

We examine the coverage properties of the predictive set constructed above under the idealized population-level
assumption that the relevant conditional quantiles are known exactly.
This analysis serves to clarify how componentwise miscoverage levels propagate to a tensor-valued predictive set and
motivates the subsequent finite-sample conformal calibration procedure.

For each component $(i,J)$, consider the scalar random variable $P_{iJ}$ and define the lower and upper conditional
quantiles
\begin{equation}
\ell_{iJ}(\boldsymbol{F})
=
q_{\alpha_{iJ}^{\mathrm{lo}}}(\boldsymbol{F}),
\qquad
u_{iJ}(\boldsymbol{F})
=
q_{\alpha_{iJ}^{\mathrm{hi}}}(\boldsymbol{F}),
\end{equation}
with
\begin{equation}
\alpha_{iJ}^{\mathrm{lo}} = \frac{1}{2}\alpha_{iJ},
\qquad
\alpha_{iJ}^{\mathrm{hi}} = 1 - \frac{1}{2}\alpha_{iJ},
\end{equation}
where $\alpha_{iJ} \in (0,1)$ denotes the nominal miscoverage allocated to component $(i,J)$.

By the definition of conditional quantiles, at the population level we have
\begin{equation}
\mathbb{P}\!\left(
P_{iJ} < \ell_{iJ}(\boldsymbol{F})
\,\middle|\,
\boldsymbol{F}
\right)
=
\alpha_{iJ}^{\mathrm{lo}},
\qquad
\mathbb{P}\!\left(
P_{iJ} > u_{iJ}(\boldsymbol{F})
\,\middle|\,
\boldsymbol{F}
\right)
=
\alpha_{iJ}^{\mathrm{lo}},
\end{equation}
and consequently
\begin{equation}
\mathbb{P}\!\left(
P_{iJ} \notin [\ell_{iJ}(\boldsymbol{F}), u_{iJ}(\boldsymbol{F})]
\,\middle|\,
\boldsymbol{F}
\right)
=
\alpha_{iJ}.
\label{eq:component_miscoverage}
\end{equation}

The tensor-valued predictive set $\mathcal{U}(\boldsymbol{F})$ is defined as the Cartesian product of the
componentwise intervals,
\begin{equation}
\mathcal{U}(\boldsymbol{F})
=
\left\{
\boldsymbol{P} \in \mathbb{R}^{3\times 3}
:\;
\ell_{iJ}(\boldsymbol{F}) \le P_{iJ} \le u_{iJ}(\boldsymbol{F})
\ \text{for all } i,J
\right\}.
\end{equation}
Equivalently,
\begin{equation}
\boldsymbol{P} \notin \mathcal{U}(\boldsymbol{F})
\quad\Longleftrightarrow\quad
\bigcup_{i,J}
\left\{
P_{iJ} \notin [\ell_{iJ}(\boldsymbol{F}), u_{iJ}(\boldsymbol{F})]
\right\}.
\end{equation}

Applying the union bound and using~\eqref{eq:component_miscoverage}, we obtain the population-level bound
\begin{align}
\mathbb{P}\!\left(
\boldsymbol{P} \notin \mathcal{U}(\boldsymbol{F})
\,\middle|\,
\boldsymbol{F}
\right)
&\le
\sum_{i,J}
\mathbb{P}\!\left(
P_{iJ} \notin [\ell_{iJ}(\boldsymbol{F}), u_{iJ}(\boldsymbol{F})]
\,\middle|\,
\boldsymbol{F}
\right)
\\[4pt]
&=
\sum_{i,J} \alpha_{iJ}.
\end{align}
In particular, if the componentwise miscoverage levels are chosen uniformly as
$
\alpha_{iJ} = \alpha/9,
$
then the conditional miscoverage of the tensor-valued set is bounded above by $\alpha$.
This bound holds without any assumptions on the dependence structure among stress components and provides a
conservative control of the overall miscoverage.

In finite-data settings, the true conditional quantiles are unknown and must be estimated.
As a result, the population-level guarantees derived above do not directly translate to finite-sample coverage.
In the following subsection, we show how conformalized quantile regression restores finite-sample, distribution-free
\emph{marginal} coverage for each stress component, and consequently for the full tensor-valued predictive set.

\section{Conformal Marginal Coverage Guarantee}
\label{appx:proof-cqr}
The componentwise conformal intervals satisfy the finite-sample guarantee due to the rank-based construction of the
conformal quantile $Q_{1-\alpha_{iJ}}(e_{iJ},\mathcal{D}_{\mathrm{cal}})$.
For a fixed component $(i,J)$, define the conformity scores
\[
e^{(n)}_{iJ}
=
\max\!\big(
\hat{q}_{\alpha^{\mathrm{lo}}_{iJ}}(\boldsymbol{F}^{(n)}) - P^{(n)}_{iJ},
\;
P^{(n)}_{iJ} - \hat{q}_{\alpha^{\mathrm{hi}}_{iJ}}(\boldsymbol{F}^{(n)})
\big),
\qquad
(\boldsymbol{F}^{(n)},\boldsymbol{P}^{(n)}) \in \mathcal{D}_{\mathrm{cal}}.
\]
Under exchangeability, the conformity score of a new test point is equally likely to take any rank among the
$|\mathcal{D}_{\mathrm{cal}}|+1$ scores
\[
\big\{
e^{(1)}_{iJ},\dots,e^{(|\mathcal{D}_{\mathrm{cal}}|)}_{iJ},
e^{\mathrm{test}}_{iJ}
\big\}.
\]
By construction,
$Q_{1-\alpha_{iJ}}$ is chosen as the
$(1-\alpha_{iJ})(1+1/|\mathcal{D}_{\mathrm{cal}}|)$–quantile of the calibration scores, implying
\[
\mathbb{P}\!\left(
e^{\mathrm{test}}_{iJ}
\le
Q_{1-\alpha_{iJ}}
\right)
\ge
1-\alpha_{iJ}.
\]
This event is equivalent to
$P_{iJ}^{\mathrm{test}} \in C_{iJ}(\boldsymbol{F}^{\mathrm{test}})$,
which establishes finite-sample, distribution-free, componentwise marginal coverage.

\section{{\color{Rtwo}Loading Condition}}
\label{appx:loading_conditions}
This appendix summarizes the three canonical incompressible loading modes used to generate and evaluate the constitutive responses: uniaxial tension, equibiaxial and biaxial tension, and pure shear. In all cases we use the deformation gradient $\boldsymbol{F}$ and enforce incompressibility via $\det(\boldsymbol{F})=1$ (equivalently $J=1$), with the associated pressure term handled separately in the stress computation.

\paragraph{Uniaxial.}
Uniaxial loading along direction 1 is prescribed by the principal stretches $
\lambda_1=\lambda,
\lambda_2=\lambda_3=\lambda^{-1/2},
$
which yields the deformation gradient $
\boldsymbol{F}_{\mathrm{uni}}(\lambda)=\mathrm{diag}\!\left(\lambda,\lambda^{-1/2},\lambda^{-1/2}\right).
$

\paragraph{Biaxial.}
Biaxial loading in directions 1 and 2 is prescribed by $
\lambda_1, \lambda_2,
\lambda_3=(\lambda_1 \lambda_2)^{-1},
$
leading to
$
\boldsymbol{F}_{\mathrm{bi}}(\lambda)=\mathrm{diag}\!\left(\lambda_1,\lambda_2,(\lambda_1 \lambda_2)^{-1}\right).
$

\paragraph{Pure shear.}
Pure shear is prescribed by opposing stretches in two directions,
$
\lambda_1=\lambda, 
\lambda_2=\lambda^{-1},
\lambda_3=1,
$
and hence
$
\boldsymbol{F}_{\mathrm{ps}}(\lambda)=\mathrm{diag}\!\left(\lambda,\lambda^{-1},1\right).
$

%\paragraph{Remarks.}The scalar loading parameter $\lambda>0$ controls the deformation magnitude and is swept over the specified range in each experiment. The corresponding stress responses are evaluated from the constitutive model for each $\mathbf{F}(\lambda)$, with the incompressibility contribution (pressure term) incorporated as described in the main text.

\section{Problem Setting}
\label{appx:params-all}

\subsection{{\color{Rtwo}Mooney}-Rivlin}
\label{appx:pr-MR}
The lower and upper quantile models employ the same monotone neural network architecture to parameterize the gradient functions, consisting of two hidden layers with 16 neurons per layer and \texttt{ReLU} activation functions. These networks are trained jointly using the Adam optimizer with a learning rate of $10^{-3}$. For quantile regression, a nominal miscoverage level of $\alpha = 0.1$ is prescribed.

\subsection{Arterial Wall Material}
\label{appx:hoz-data}
Both the lower and upper quantile models use identical network architectures with two hidden layers of width 16 and \texttt{ELU} activation functions. The two models are trained jointly using the Adam optimizer with a learning rate of $10^{-3}$. A nominal miscoverage level of $\alpha = 0.1$ is prescribed.

\subsection{Porcine Atrioventricular Valve Leaflets}
\label{appx:porcine-data}
\begin{figure}[h]
  \centering
  \begin{subfigure}[b]{0.4\textwidth}
    \includegraphics[width=\textwidth]{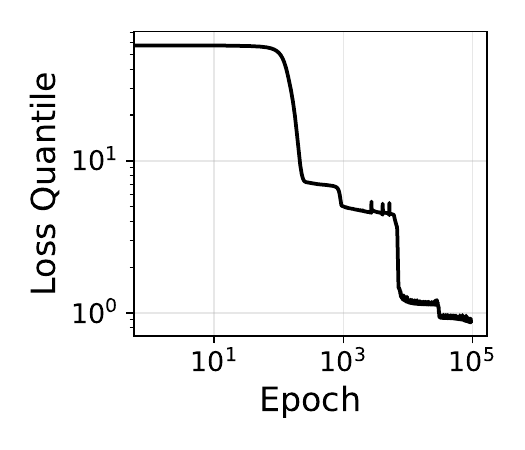}
    %\caption{}
    %\label{subfig:one}
  \end{subfigure}
    \caption{
    Pinball loss training history for the porcine atrioventricular {\color{Rtwo}valve} leaflets data.
    }
  \label{fig:pr3-loss}
\end{figure}

Both the lower and upper quantile models use identical network architectures with two hidden layers of width 16 and \texttt{ELU} activation functions. The models are trained jointly using the Adam optimizer with a learning rate of $10^{-3}$. A nominal miscoverage level of $\alpha = 0.06$ is prescribed. The training loss function is plotted in Figure \ref{fig:pr3-loss}.

\bibliographystyle{unsrt}
\bibliography{ref}

\end{document}